%% file: IEEE.tex
\documentclass[preprint]{elsarticle}

\graphicspath{{figures/}}
\DeclareGraphicsExtensions{.pdf,.jpeg,.png}

\usepackage[caption=false,font=footnotesize]{subfig}

\usepackage{algorithm}
\usepackage{algorithmic}
\algsetup{
linenosize=\small,
linenodelimiter=.
}

\usepackage[cmex10]{amsmath}
\usepackage{commath}
\usepackage{amsfonts}
\usepackage{amssymb}

\usepackage[mathscr]{euscript}

\usepackage{bm}

\usepackage{url}

\usepackage{lineno,hyperref}
\modulolinenumbers[5]

\usepackage{multirow}

\usepackage{color}
\usepackage{colortbl}
\definecolor{Gray_my}{gray}{0.9}

\newcommand{\responsess}[1]{\textnormal{#1}}

\definecolor{gray}{rgb}{0.5,0.5,0.5}

\usepackage{siunitx}
\usepackage{enumerate}

\usepackage{amsthm}
\newtheorem{thm}{Theorem}

\newdefinition{defs}{Definition}

\usepackage{natbib}

\journal{Elsevier}

\begin{document}

\begin{frontmatter}


\title{ Multi-Slice Low-Rank Tensor Decomposition Based Multi-Atlas Segmentation: Application to Automatic Pathological Liver CT Segmentation }



\author[mymainaddressnew,mymainaddress]{Changfa~Shi}

\address[mymainaddressnew]{Mobile E-business Collaborative Innovation Center of Hunan Province, Hunan University of Technology and Business, Changsha 410205, China}
\address[mymainaddress]{Department of Computer Science, Utah State University, Logan, UT 84322, USA}

\author[mymainaddress2]{Min~Xian\corref{mycorrespondingauthor}}
\cortext[mycorrespondingauthor]{Corresponding author. Tel: +1-208-757-5425.}
\ead{mxian@uidaho.edu}

\author[mymainaddressnew]{Xiancheng~Zhou}
\author[mymainaddress2]{Haotian~Wang}

\author[mymainaddress]{Heng-Da~Cheng}
 
\address[mymainaddress2]{Department of Computer Science, University of Idaho, Idaho Falls, ID 83402, USA}


\input{chapters/abstract}

\end{frontmatter}
\clearpage

\input{chapters/introduction}
\input{chapters/relatedwork}
\input{chapters/LRTD}
\input{chapters/methods}
\input{chapters/results}
\input{chapters/conclusion}

%
%

\section*{Acknowledgment}


%
This work was supported in part by 
the National Natural Science Foundation of China under Grant No. 61701178, 
and the Natural Science Foundation of Hunan Province of China (No. 2018JJ3256),
and the China Scholarship Council (No. 201908430083).


\appendix

\input{chapters/appendix}

\section*{References}











\bibliographystyle{reference/model2-names.bst}\biboptions{authoryear}
\bibliography{reference/IEEEabrv,reference/IEEEexample}



\end{document}

%% file: chapters/abstract.tex
\begin{abstract}

Liver segmentation from abdominal CT images is an essential step for liver cancer computer-aided diagnosis and surgical planning. However, both the accuracy and robustness of existing liver segmentation methods cannot meet the requirements of clinical applications. In particular, for the common clinical cases where the liver tissue contains major pathology, current segmentation methods show poor performance. In this paper, we propose a novel low-rank tensor decomposition (LRTD) based multi-atlas segmentation (MAS) framework that achieves accurate and robust pathological liver segmentation of CT images. Firstly, we propose a multi-slice LRTD scheme to recover the underlying low-rank structure embedded in 3D medical images. It performs the LRTD on small image segments consisting of multiple consecutive image slices. Then, we present an LRTD-based atlas construction method to generate tumor-free liver atlases that mitigates the performance degradation of liver segmentation due to the presence of tumors. Finally, we introduce an LRTD-based MAS algorithm to derive patient-specific liver atlases for each test image, and to achieve accurate pairwise image registration and label propagation. Extensive experiments on three public databases of pathological liver cases validate the effectiveness of the proposed method. Both qualitative and quantitative results demonstrate that, in the presence of major pathology, the proposed method is more accurate and robust than state-of-the-art methods.

\end{abstract}

\begin{keyword}
Pathological liver segmentation, Low-rank tensor decomposition, Multi-atlas segmentation, Tensor robust PCA, $\star_{M}$-product.
\end{keyword}

%% file: chapters/introduction.tex
\section{Introduction}

According to GLOBOCAN 2018 estimates, 
liver cancer is the sixth most common cancer 
and the fourth leading cause of cancer mortality around the world~\citep{Bray18}.
Liver segmentation 
(the extraction of the liver from its surrounding tissue)
of abdominal computed tomography (CT) images 
is a key step and a prerequisite for 
liver cancer computer-aided diagnosis (CAD), 
surgical planning and other interventional procedures. 
However,
in current clinical practice,
liver segmentation from abdominal CT images 
is still predominantly performed 
manually by expert radiologists
in a slice-by-slice manner. 
Due to the abundance of the CT images for each patient,
the manual segmentation is time-consuming,
and subject to observer error and personal bias.
Therefore, 
it is highly desirable to develop
fully automated liver segmentation approaches
that can efficiently and automatically 
extract the liver boundary without any user intervention.

Numerous liver segmentation methods have been published
in the last few decades.
They can be generally classified into 
traditional image-based methods, model-based methods, 
and deep learning-based methods~\citep{Erdt12}.
The traditional image-based methods are
mainly relies on image intensity information
to perform liver segmentation, 
such as intensity thresholding~\citep{Kobashi95}
and region growing~\citep{Rusko09},
and they tend to have poor performance for clinical liver cases.
The model-based methods,
such as active shape model (ASM)~\citep{Heimann09_Review} 
and multi-atlas segmentation (MAS)~\citep{Iglesias15},
have yielded remarkable results in liver CT segmentation~\citep{Heimann09},
where shape, appearance and spatial location information of the liver tissue
were incorporated into the segmentation framework as prior knowledge.
In recent years,
deep learning-based methods achieve popularity
in the field of medical image analysis
due to their tremendous success in the computer vision community, 
and they have achieved state-of-the-art performance 
in medical image segmentation~\citep{Litjens17}.
The main advantage of deep learning-based methods 
is that the most relevant features are automatically generated
and selected for the given problem,
rather than being manually engineered.
%
However,
when applied to liver CT image segmentation,
the model-based methods proved to have
comparable performance to deep learning-based methods~\citep{Ahn19}.
%
The main reason is that
the performance of deep learning-based methods
highly depends on the availability of massive amounts of training data,
which cannot be fully met in the case of liver CT image segmentation.
Furthermore,
compared to model-based methods,
the interpretability of  
deep learning-based methods is poor, 
which, however, is of paramount importance in clinical applications.

Nevertheless, 
the drawback of
aforementioned liver segmentation methods 
is that they still cannot meet the performance requirements of 
clinical applications.
In particular, for common clinical cases 
where the liver tissue contains major pathology, 
current liver segmentation methods still show poor performance.
It is mainly because the hypotheses of
most current liver segmentation methods  
are only applicable to segment liver tissue 
of healthy or minor pathology conditions,
rather than of major pathology.
%

\begin{figure*}[!t]
	{
		\begin{center}
			\hfill
			\subfloat[]
			{
				\includegraphics[width=0.31\textwidth]{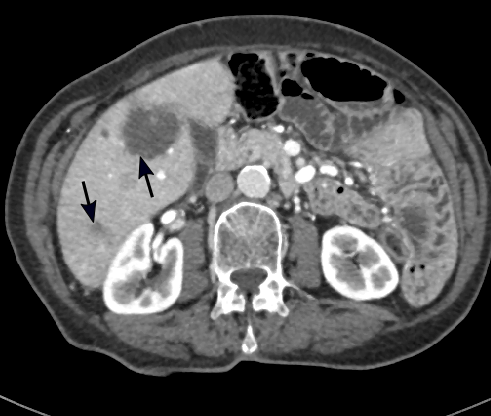}
				\label{fig:TumorChallenges1}
			}
			\hfill
			\subfloat[]
			{
				\includegraphics[width=0.31\textwidth]{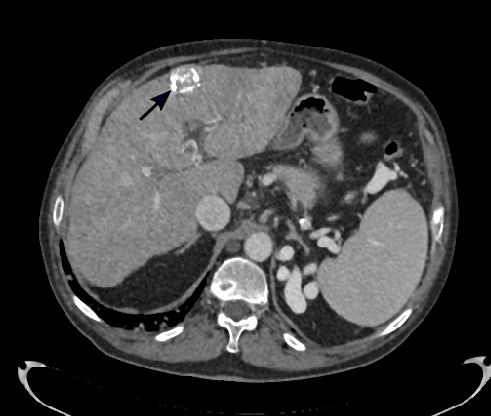}
				\label{fig:TumorChallenges2}
			}
			\hfill
			\subfloat[]
			{
				\includegraphics[width=0.31\textwidth]{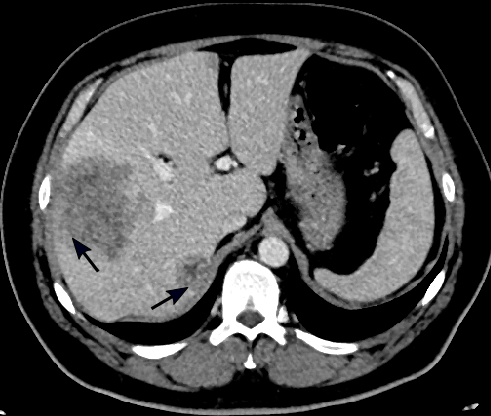}
				\label{fig:TumorChallenges3}
			}
		\end{center}
	}
	\caption{ Examples demonstrating challenges in accurate and robust 
			  pathological liver segmentation in CT scans,
			  consisting of liver tissue with 
			  (a) hypodense tumor, (b) hyperdense tumor,
			  and (c) tumors located at liver boundary.
			}
	\label{fig:TumorChallenges}
\end{figure*}

Specifically, 
the first challenge of pathological liver segmentation
is the presence of large tumors,
which exhibit totally different intensity values 
from that of the normal liver tissue.
The large tumors can cause
various undesired shape deformation of the liver tissue.
Moreover, the tumors show very large variability of size and image appearance,
resulting in tumors with different contrast levels, 
such as hypodense tumors (Fig. \ref{fig:TumorChallenges1}) and 
hyperdense tumors (Fig. \ref{fig:TumorChallenges2}). 
The second challenge of pathological liver segmentation
is the complex spatial variability of tumors
at liver boundary or inside liver parenchyma.
In particular, 
when the tumors are at liver boundary 
(Fig. \ref{fig:TumorChallenges3}),
there is a very high probability that
the peripheral liver tissue  
will be excluded from the final segmentation results, 
leading to under-segmentation.
It is mainly due to the existence of blurred boundaries 
between the peripheral tumors and 
the nearby tissue (e.g., the muscle tissue)
exhibiting similar image appearance. 
%
%
%

Therefore, in the routine clinical setting,
pathological liver segmentation
is still mostly performed manually by expert radiologists,
which is very labor-intensive,
and subject to high intra- and inter-operator variability.
Furthermore,
since the liver tissue with large tumors exhibits
totally different shape and image appearance
from that of the normal liver tissue,
it is also very challenging even
for the expert radiologists to perform the segmentation task manually.
Recently,
a few pathological liver segmentation methods
have been proposed in the literature   
(refer to Section~\ref{subsec:Pathological_Review} 
for a comprehensive survey).
Nevertheless,
the performance of these segmentation approaches
is still unsatisfactory
in the presence of major pathology.

In this paper, 
to address the above-mentioned issues,
we integrate the general $\star_{M}$-product based 
low-rank tensor decomposition (LRTD) theory into 
the widely used MAS framework, 
and propose a novel automatic method for 
accurate and robust pathological liver segmentation 
of abdominal CT images.
It is well-known that 
the segmentation accuracy of the MAS framework highly depends on
the quality of the constructed atlases and 
pairwise image registrations~\citep{Iglesias15}.
When employed to perform pathological liver CT segmentation,
the main challenge to traditional MAS framework
is the presence of large tumors, 
which exhibit totally different intensity values 
from that of the normal liver tissue, 
leading to liver atlases of low quality
and large errors in pairwise image registrations.
%
%
Inspired by the recently popular LRTD theory~\citep{Kolda11},
also known as Tensor Robust Principal Component Analysis (TRPCA)~\citep{Lu20}, 
in the fields of signal processing and computer vision,
we first propose a multi-slice LRTD scheme 
for low-rank structure learning in 3D medical images.
Specifically, 
we partition each liver CT image 
into smaller segments consisting of multiple consecutive image slices, 
and perform the LRTD on each segment sequentially.
Then we present an LRTD-based atlas construction method 
to obtain tumor-free liver atlases  
that mitigate the performance degradation of liver segmentation 
caused by the presence of tumors.
Specifically,
given a data tensor $\bm{\mathcal{D}}$, 
the LRTD model decomposes it into two parts:
(1) a low-rank component $\bm{\mathcal{L}}$ 
corresponding to the tumor-free liver images via tensor rank minimization,
and (2) a sparse component $\bm{\mathcal{E}}$
corresponding to the sparse tumors via $\ell_0$-norm minimization.
Thus, the LRTD model fits the goal 
of deriving tumor-free liver images very well. 
Furthermore,
we introduce an LRTD-based MAS algorithm 
to derive patient-specific liver atlases for each test image,
and to yield accurate pairwise image registration and label propagation.

In order to evaluate the performance of our proposed MAS-based segmentation framework,
and to show its clinical applicability to pathological liver segmentation,
we extensively tested it using 
three public clinical CT databases,
and also compared it with state-of-the-art methods.
%
The experimental results
demonstrate that the proposed method 
yields higher accuracy and robustness than that of 
state-of-the-art methods.
%
%
%
%
%
%


The main contributions of the proposed LRTD-based MAS framework
can be summarized as follows:
\begin{enumerate}[(1)] 
  
  \item A general multi-slice LRTD scheme is proposed to recover the underlying low-rank structure embedded in 3D medical images. 
  In particular, the discrete cosine transform (DCT) converts the calculation of tensor singular value decomposition (t-SVD) to matrix SVD in the transform domain, which enhances the computational efficiency and maintains spatial relationship. 
  Then the tensor singular value thresholding (t-SVT) algorithm is used to recover the underlying low-rank structure. 
  The general scheme is also applicable to other medical imaging modalities and organs (~\cite{Qin19,Xian18,Khaleel18})
  (Section~\ref{subsec:LRTD-PAs}).
  
  \item  An LRTD-based atlas construction method is developed to produce tumor-free liver atlases. In the MAS framework, atlases with tumor regions will lead to inaccurate pairwise image registration and performance degradation
  (Section~\ref{subsec:LRTD-PAs}).
  
  \item An LRTD-based MAS algorithm is proposed to achieve
  more accurate and robust liver segmentation. 
  An atlas selection strategy is first implemented
  to obtain patient-specific liver atlases for each test image.
  Then, based on the selected tumor-free liver atlases, it generates a tumor-free test image that yields accurate pairwise image registration and label propagation (Section~\ref{subsec:LRTD-MAS}).
  
   \item We conducted extensive experiments to compare the proposed method to state-of-the-art methods using three public clinical CT databases. It is shown that our method is more accurate and robust than state-of-the-art methods in the presence of major pathology (Section~\ref{section:Results}). 
\end{enumerate}
%

%

%% file: chapters/relatedwork.tex
\section{ Related Work }



\subsection{Pathological Liver CT Segmentation}
\label{subsec:Pathological_Review}



~\citet{Li20}
proposed a pathological liver CT segmentation method
by combining level set, sparse shape composition, and graph cut methods.
The initial liver shape was obtained by a level set method 
integrated with intensity bias and position constraint,
followed by sparse shape composition and graph cut 
based segmentation refinements.
~\citet{Raju20}
proposed a user-guided domain adaptation framework
for pathological liver CT segmentation,
which used prediction-based adversarial domain adaptation  
to guide mask predictions by the user interactions.
%
%
%
~\citet{Dakua16}
proposed a semi-automatic method
for pathological liver CT segmentation,
where a stochastic resonance algorithm was 
utilized to enhance the contrast of
the liver images,
followed by cellular automaton 
and level set segmentation methods.
%
%
\citet{Umetsu14} proposed a graph cut based method 
for segmenting liver CT cases with unusual shapes 
and pathological lesions,
where a sparse representation based 
patient-specific probabilistic atlas (PA) reinforced by the lesion bases 
was incorporated with the graph cut method.
Nevertheless,
in the presence of major pathology,
the performance of
the above-mentioned liver CT segmentation methods
is still not satisfactory,
and further improvements are needed.

\subsection{\responsess{Atlas-based Approaches for Liver CT Segmentation}}
\label{subsec:Pathological_Review}

\responsess{
The atlas-based method is one of the most popular model-based methods 
for liver segmentation~\citep{Heimann09},
where the shape prior knowledge is represented by liver atlases. 
Atlas-based methods are mainly in three categories~\citep{Iglesias15}:
single atlas segmentation, PA segmentation~\citep{Park03},
and MAS~\citep{Rohlfing04}.
}

\responsess{
~\citet{Park03} first proposed a Bayesian framework based PA method 
to segment liver and other abdominal organs in CT images, 
where an image registration method based on the thin-plate spline 
was employed to construct the PAs. 
%
%
~\citet{Slagmolen07_Workshop} and ~\citet{Okada08} 
also used PA method to perform liver CT segmentation. 
When constructing liver atlases, 
~\citet{Slagmolen07_Workshop} introduced a new surface distance penalty 
to minimize the distance between the segmentations 
on reference and floating images. 
While spatial normalization using the abdominal cavity 
was first performed in~\citet{Okada08} before the construction of liver atlases.
~\citet{Rikxoort07_Workshop}, ~\citet{Linguraru10}, and ~\citet{Platero14}
employed the PA method to perform initial liver CT segmentation, 
and then
used voxel classification, geodesic active contour, 
and MAS for segmentation refinement, respectively.
~\citet{Xu16SIFT} proposed an MAS approach for liver CT segmentation 
using 3D scale-invariant feature transform (SIFT) flow, 
where a 3D-based nonparametric label transfer method 
was introduced to merge the labels of registered atlases based on SIFT features.
}

\responsess{
Although existing atlas-based methods have achieved high performance 
on liver CT segmentation, 
their performance deteriorates significantly
when segmenting liver tissue with major pathology~\citep{Platero14}.
}







\subsection{Low-Rank Tensor Decomposition for Medical Image Computing}
The LRTD theory 
has been widely studied and applied in 
the fields of signal processing and computer vision~\citep{Sidiropoulos17, Cichocki15, Sobral17}.
Recently, it
has gained wide applications in the 
field of medical image computing~\citep{Madathil19},
including medical image reconstruction, super-resolution, 
denoising, and analysis.
The images from most medical imaging modalities
are currently acquired as 3D volumes,
such as CT, Magnetic Resonance (MR).
Since the LRTD models the 3D volumes
in their native format (i.e., tensor)
rather than vectorizing them into 1D vectors,
it can simultaneously exploits 
both the spatial and temporal correlations
embedded in the 3D volumes~\citep{Madathil19}.
%
%

~\cite{Liu20} and~\cite{Roohi17}
proposed multi-dimensional approaches
to the problem of dynamic MRI reconstruction  
of under-sampled k-space
by formulating it as an LRTD problem,
and the recovery performance was superior to 
that of other reconstruction methods.
%
\responsess{~\cite{Wu18}
proposed an improved tensor dictionary learning method, 
called $\ell_0$TDL, 
for low-dose spectral CT reconstruction 
with a constraint of image gradient $\ell_0$-norm.}
~\cite{Shi15}
proposed an MR image super-resolution method that
integrated both local and global information for effective image
recovery via total variation and low-rank tensor regularizations,
respectively.
%
\responsess{~\cite{Hatvani19}
proposed a tensor-factorization-based method 
for 3D single image super-resolution of Dental CT.
}
A few LRTD-based image denoising methods
were proposed for MR~\citep{Khaleel18, Fu16} 
and CT images~\citep{Sagheer19}
to reduce noise and artifacts introduced 
during image acquisition. 
~\cite{Jiang20}
proposed a functional connectivity network estimation approach 
by the assumption that the functional connectivity networks
have similar topology across subjects via the LRTD.
~\cite{Qin19}
proposed an LRTD-based method for 
accurately recovering vessel structures and intensity information
from the X-ray coronary angiography (XCA) sequences.
~\cite{Xu16}
proposed an LRTD-based one-step method 
for axial alignment in 360-degree anterior chamber optical coherence tomography.

\responsess{
Furthermore,
~\cite{Lu16} first proposed the t-product based TRPCA framework
for face recovery and image denoising problems,
and established a theoretical bound for the exact recovery
under certain tensor incoherence conditions.
The proposed approach differs from 
existing LRTD-based approaches in two important ways.
\begin{enumerate}[(1)] 
    \item Methodology. 
	Existing LRTD-based methods used 
	the classical canonical polyadic decomposition (CPD)~\citep{Kolda09}, 
	Tucker decomposition~\citep{Kolda09} 
	or t-product based tensor SVD (t-SVD)~\citep{Lu16} 
	to define tensor nuclear norm, 
	while we employ the new $\star_{M}$-product based t-SVD~\citep{Kilmer19}.
	The CPD is generally NP-hard to compute, 
	while the Tucker decomposition can be substantially suboptimal~\citep{Lu16}.
	Moreover, the superiority of the $\star_{M}$-product over the t-product 
	is well demonstrated in ~\cite{Kilmer19}.	
	Furthermore, we propose a novel multi-slice LRTD scheme
	to learn low-rank structure embedded in 3D medical images. 
    \item Application. 
	Existing LRTD-based methods were mainly applied for medical image reconstruction, 
	super-resolution, and denoising. 
	To the best of our knowledge, 
	the proposed method is the first MAS framework directly 
	utilizing the LRTD theory for medical image segmentation.	
\end{enumerate}
}
%
%

\subsection{Relation to Previous Work}

The proposed LRTD-based MAS framework is partly inspired by 
two recent papers (~\citet{Liu15} and ~\citet{shih17_MIA}).
They both proposed an atlas-based organ segmentation method
by utilizing low-rank matrix decomposition (LRMD) theory~\citep{Zhou-Review14} 
to handle clinical cases with pathology.
In this study, 
we significantly extend their methods in the following directions:
\begin{enumerate}[(1)] 
\item We propose a new MAS framework based on the LRTD,
rather than the LRMD as in their original methods.
To perform the LRMD,
the 3D data tensor is first reformatted to a matrix
by vectorizing voxel intensity values of each CT scan 
to form the column vectors.
The local spatial information is thus completely lost,
and the multi-dimensional structure 
embedded in the tensor data is disregarded,
leading to considerable performance degradation.
While our tensor-based method 
can fully exploit 
the intrinsic three-dimensional structural information of the CT scans,
resulting in much more accurate data decomposition results.
\item The whole image volumes 
were directly employed to perform the LRMD in their original methods,
whereas we propose a new multi-slice LRTD scheme,
where each liver CT image is first partitioned into  
smaller segments consisting of multiple consecutive image slices, 
then the LRTD is performed on each segment sequentially.
Since smaller image segment exhibits fewer overall structural changes,
it will lie on a low-rank subspace~\citep{Lee18}.
It can thus yield more accurate data decomposition results than 
directly performing the LRMD on whole image volumes.
\end{enumerate}
 

%% file: chapters/LRTD.tex
\section{ Mathematical Notations and Tensor Preliminaries }
\label{subsec:tensor-math}


In this section, 
the notations and preliminaries of tensor 
used in the rest of the paper are briefly described.
Throughout this paper, 
we follow the notation and terminology of
~\cite{Kolda09}, ~\cite{Lu19} and~\cite{Kilmer19}.

\subsection{ Mathematical Notations }
A third-order tensor and its entries are denoted by 
capital boldface Euler script letters and small symbols, respectively, 
e.g., $x_{ijk}$ is the $(i, j, k)$-th entry of 
the tensor $\bm{\mathcal{X}} \in \mathbb{R}^{n_1 \times n_2 \times n_3}$.
Matrix is denoted by capital boldface letters, e.g., $\mathbf{X} \in \mathbb{R}^{n_1 \times n_2}$.
The two-dimensional horizontal, lateral, and frontal slices of a third-order tensor $\bm{\mathcal{X}}$ 
are denoted by $\bm{\mathcal{X}}(i,:,:)$, $\bm{\mathcal{X}}(:,j,:)$, 
and $\bm{\mathcal{X}}(:,:,k)$, respectively. 
The frontal slices are often denoted more compactly as $\mathbf{X}^{(k)}$, 
i.e., $\mathbf{X}^{(k)}=\bm{\mathcal{X}}(:,:,k)$.  
The mode-$k$ fibers of a third-order tensor $\bm{\mathcal{X}}$ 
are vectors defined by fixing all indices but the $k$-th, e.g., the mode-3 fibers are denoted by $\bm{\mathcal{X}}(i,j,:)$. 
 
The inner product between two matrices is defined as $\left \langle \mathbf{X}, \mathbf{Y} \right \rangle = \texttt{tr}(\mathbf{X}^{*}\mathbf{Y}) $, where $\mathbf{X}^{*}$ and $\texttt{tr}(\cdot)$ denote the conjugate transpose of $\mathbf{X}$ and the matrix trace, respectively. 
The inner product between two tensors can then be defined as $\left \langle \bm{\mathcal{X}}, \bm{\mathcal{Y}} \right \rangle = \sum_{k=1}^{n_3} \left \langle \mathbf{X}^{(k)}, \mathbf{Y}^{(k)} \right \rangle$.
Three tensor norms are used: 
$\|\bm{\mathcal{X}}\|_0$ the $\ell_0$-norm 
(i.e., the number of non-zero entries in $\bm{\mathcal{X}}$), 
$\|\bm{\mathcal{X}}\|_1 = \sum_{ijk}  { |x_{ijk}| }$ the $\ell_1$-norm,
and $\|\bm{\mathcal{X}}\|_{\infty} = \operatorname*{max}_{ijk} |x_{ijk}|$ the infinity norm.

\subsection{ Tensor Preliminaries }
\label{subsec:tensor-preliminaries}

\textbf{Mode-3 product}.
The mode-3 product ($\times_{3}$) is an operator between a tensor $\bm{\mathcal{X}} \in \mathbb{R}^{n_1 \times n_2 \times n_3}$ and a matrix $\mathbf{M} \in \mathbb{R}^{n_3 \times n_3}$. The result is another tensor $\bar{\bm{\mathcal{X}}} \in \mathbb{R}^{n_1 \times n_2 \times n_3}$ defined as the following~\citep{Kolda09}:
\begin{equation}
		\bar{\bm{\mathcal{X}}} = \bm{\mathcal{X}} \times_{3} \mathbf{M} = \mathbf{M} \cdot \bm{\mathcal{X}}_{(3)},
\end{equation}
where $\bm{\mathcal{X}}_{(3)} \in \mathbb{R}^{n_3 \times{n_1n_2}}$ 
denotes the mode-3 unfolding of $\bm{\mathcal{X}}$, 
defined as a matrix whose columns consist of the mode-3 fibers. 
The mode-3 product can be interpreted geometrically as performing a linear transform 
on $\bm{\mathcal{X}}$ along the third dimension via transform matrix $\mathbf{M}$.
We can denote it 
more compactly as $\mathbf{M}(\bm{\mathcal{X}})$.

\textbf{t-product}.
The t-product ($*$) between two tensors
$\bm{\mathcal{X}} \in \mathbb{R}^{n_1 \times n_2 \times n_3}$
and $\bm{\mathcal{Y}} \in \mathbb{R}^{n_2 \times m \times n_3}$
is a multiplication operator
that preserves the order of the tensor.
The result tensor
$\bm{\mathcal{Z}} \in \mathbb{R}^{n_1 \times m \times n_3}$ is
defined as following~\citep{Kilmer11}:
\begin{equation}
		\bm{\mathcal{Z}} = \bm{\mathcal{X}}*\bm{\mathcal{Y}} = \texttt{fold} \left( \texttt{bcirc}(\bm{\mathcal{X}}) \cdot \texttt{unfold}(\bm{\mathcal{Y}}) \right),
\end{equation}
where $\texttt{bcirc}(\bm{\mathcal{X}}) \in \mathbb{R}^{n_1n_3 \times n_2n_3}$ is a block circulant matrix,
which can be regarded as a new matricization of $\bm{\mathcal{X}}$,
\texttt{fold} and \texttt{unfold} are a pair of operators on tensors 
(see Eq. \ref{eq:LRSD-PCP1} and Eq. \ref{eq:LRSD-PCP2} 
in \ref{sec_Appen:T-SVD}).

For tensor $\bm{\mathcal{X}} \in \mathbb{R}^{n_1 \times n_2 \times n_3}$, 
its discrete Fourier transform (DFT) along the third dimension is denoted as  
$\bar{\bm{\mathcal{X}}} = \bm{\mathcal{X}} \times_{3} \mathbf{F}_{n_3} = \mathbf{F}_{n_3}(\bm{\mathcal{X}})$, where $\mathbf{F}_{n_3} \in \mathbb{C}^{n_3 \times n_3}$ is the DFT matrix. By using Matlab convention, we also have $\bar{\bm{\mathcal{X}}} = \texttt{fft}(\bm{\mathcal{X}}, [ ], 3)$ .
Conversely, $\bm{\mathcal{X}}$ can be derived from $\bar{\bm{\mathcal{X}}}$ via inverse DFT,  
i.e., $\bm{\mathcal{X}} = \bm{\mathcal{X}} \times_{3} \mathbf{F}^{-1}_{n_3} = \mathbf{F}^{-1}_{n_3}(\bm{\mathcal{X}}) = \texttt{ifft}(\bar{\bm{\mathcal{X}}}, [], 3)$.
%
%
%
%
%
%
The block circulant matrix $\texttt{bcirc}(\bm{\mathcal{X}})$ can be block-diagonalized to 
a special block diagonal matrix $\bar{\bm{\mathbf{X}}}$,
whose main diagonal blocks are the frontal slices of $\bar{\bm{\mathcal{X}}}$
(see Eq. \ref{eq:LRSD-PCP} in \ref{sec_Appen:T-SVD}), 
via the DFT matrix $\mathbf{F}_{n_3} \in \mathbb{C}^{n_3 \times n_3}$~\citep{Kolda09}:
\begin{equation}
\label{eq:blockdft}
  (\mathbf{F}_{n_3} \otimes \mathbf{I}_{n_1}) \cdot \texttt{bcirc}(\bm{\mathcal{X}}) 
  \cdot (\mathbf{F}^{-1}_{n_3} \otimes \mathbf{I}_{n_2})= \bar{\mathbf{X}},  
\end{equation}
where $\otimes$ denotes the Kronecker product.

Based on the block-diagonalized property of $\texttt{bcirc}(\bm{\mathcal{X}})$ in Eq. \ref{eq:blockdft}, the t-product can also be defined as the matrix-matrix product in the DFT domain~\citep{Lu20}:
\begin{equation}
		  \bm{\mathcal{Z}} = (\bar{\bm{\mathcal{X}}} \bigtriangleup \bar{\bm{\mathcal{Y}}}) \times_{3} \mathbf{F}^{-1}_{n_3}  = \mathbf{F}^{-1}_{n_3} \left( \mathbf{F}_{n_3}(\bm{\mathcal{X}}) \bigtriangleup \mathbf{F}_{n_3}(\bm{\mathcal{Y}}) \right) ,  
\end{equation}
where $\bigtriangleup$ denotes the face-wise product, defined as the matrix-matrix product between corresponding frontal slices of the two tensors.

\textbf{$\star_{M}$-product}.
A new tensor-tensor product operator, 
called $\star_{M}$-product, is proposed in ~\cite{Kilmer19}.  
It can convert the data into other transform domains under any invertible matrix $\mathbf{M}$,
rather than the specific DFT domain as in the t-product.
The superiority of $\star_{M}$-product
compared to the t-product is demonstrated in ~\cite{Kilmer19}.

Let $\mathbf{M} \in \mathbb{R}^{n_3 \times n_3}$ be any invertible matrix satisfying:
\begin{equation}
\label{eq:tranM}
		\bm{\mathbf{M}}^{*} \bm{\mathbf{M}} = \bm{\mathbf{M}} \bm{\mathbf{M}}^{*} = l\bm{\mathbf{I}}_{n_3},  
\end{equation}
where $l>0$ is a constant,
$\bm{\mathcal{X}} \in \mathbb{R}^{n_1 \times n_2 \times n_3}$ and  
$\bm{\mathcal{Y}} \in \mathbb{R}^{n_2 \times m \times n_3}$ are two tensors. 
Then the $\star_{M}$-product $\bm{\mathcal{X}} \star_{M} \bm{\mathcal{Y}}$ 
results in a tensor $\bm{\mathcal{Z}} \in \mathbb{R}^{n_1 \times m \times n_3}$
defined as below~\citep{Kilmer19,Lu19}:
\begin{equation}
		 \bm{\mathcal{Z}} = \bm{\mathcal{X}}\star_{M}\bm{\mathcal{Y}} = (\bar{\bm{\mathcal{X}}} \bigtriangleup \bar{\bm{\mathcal{Y}}}) \times_{3} \mathbf{M}^{-1} = \mathbf{M}^{-1} \left( \mathbf{M}(\bm{\mathcal{X}}) \bigtriangleup \mathbf{M}(\bm{\mathcal{Y}}) \right),  
\end{equation}
where $\bar{\bm{\mathcal{X}}} = \mathbf{M}(\bm{\mathcal{X}}) = \bm{\mathcal{X}} \times_{3} \mathbf{M}$ denotes the tensor in the transform domain induced by the invertible transformation matrix $\mathbf{M}$.
Note that the t-product becomes a special case of the $\star_{M}$-product when $\mathbf{M}=\mathbf{F}_{n_3}$.

Based on the $\star_{M}$-product, 
the following main concepts of tensor can be defined:
conjugate transpose, identity tensor, orthogonal tensor, 
and f-diagonal tensor
(see \ref{sec_Appen:T-SVD}).
Also, we can obtain the following theorem
defining the t-SVD:
\begin{thm}[\textbf{T-SVD}~\cite{Kilmer19}]
\label{shrinkage}
Let $\bm{\mathcal{X}} \in \mathbb{R}^{n_1 \times n_2 \times n_3}$,
and $\mathbf{M} \in \mathbb{R}^{n_3 \times n_3}$ be any invertible matrix.
Then $\bm{\mathcal{X}}$ can be factorized as:
\begin{equation}
\label{eq:t-SVD}
		\bm{\mathcal{X}} = \bm{\mathcal{U}} \star_{M} \bm{\mathcal{S}} \star_{M} \bm{\mathcal{V}}^{*},  
\end{equation}
where $\bm{\mathcal{U}} \in \mathbb{R}^{n_1 \times n_1 \times n_3}$ and 
$\bm{\mathcal{V}} \in \mathbb{R}^{n_2 \times n_2 \times n_3}$ are orthogonal tensors, and 
$\bm{\mathcal{S}} \in \mathbb{R}^{n_1 \times n_2 \times n_3}$ is an f-diagonal tensor.
\end{thm}

An illustration of the t-SVD factorization is shown in Fig.~\ref{fig:tsvd}
in \ref{sec_Appen:T-SVD}.
\responsess{Empirically,
after converting $\bm{\mathcal{X}}$ 
into transform domain as $\bar{\bm{\mathcal{X}}}$,
we can obtain the t-SVD factorization results 
by computing matrix SVD on each frontal slice of $\bar{\bm{\mathcal{X}}}$, 
refer Algorithm~\ref{algo:tsvd} in \ref{sec_Appen:T-SVD}.
When performing the matrix SVD on each slice matrix in the spatial domain, 
one slice is considered at a time, 
thus losing the important structural information provided by neighboring slices, 
which will cause significant performance degradation of data decomposition. 
The t-SVD can fully exploit the intrinsic multi-dimensional structure 
embedded in the tensor data, 
and produce much smoother and more homogeneous low-rank liver images 
(Fig. \ref{fig:Cluster-PAss}).
}

\textbf{Tensor nuclear norm}.
The minimization of the tensor rank 
is known as NP-hard due to the non-convexity nature~\citep{Hillar13}. 
Fortunately, as in the matrix case, 
the tensor nuclear norm can be employed 
as a convex relaxation of the tensor rank
for the LRTD~\citep{Lu19}.

Let $\mathbf{M} \in \mathbb{R}^{n_3 \times n_3}$ 
be any invertible matrix satisfying Eq. \ref{eq:tranM},
the tensor nuclear norm of $\bm{\mathcal{X}} \in \mathbb{R}^{n_1 \times n_2 \times n_3}$
is defined as $\|\bm{\mathcal{X}}\|_{*} = \frac{1}{l} \sum_{i=1}^{n_3} \|\bar{\mathbf{X}}^{(i)} \|_{*}$~\citep{Lu19}.
%
%
%
%
%
It has been proven that
the tensor nuclear norm $\|\bm{\mathcal{X}}\|_{*}$ 
is the convex envelop of the tensor average rank $\operatorname*{rank_a}(\bm{\mathcal{X}})= \frac{1}{l} \sum_{i=1}^{n_3} \operatorname*{rank} ( \bar{\mathbf{X}}^{(i)} )$ 
within the unit ball of the tensor spectral norm 
$\|\bm{\mathcal{X}}\| = \|\bar{\bm{\mathbf{X}}}\|$~\citep{Lu19}.
We will thus use the tensor nuclear norm $\|\cdot\|_{*}$ 
to characterize the low-rank latent structure of a tensor.


%% file: chapters/methods.tex
\section{ Methods }



\begin{figure*}[!t]
	\begin{center}
		\includegraphics[width=1.1\textwidth]{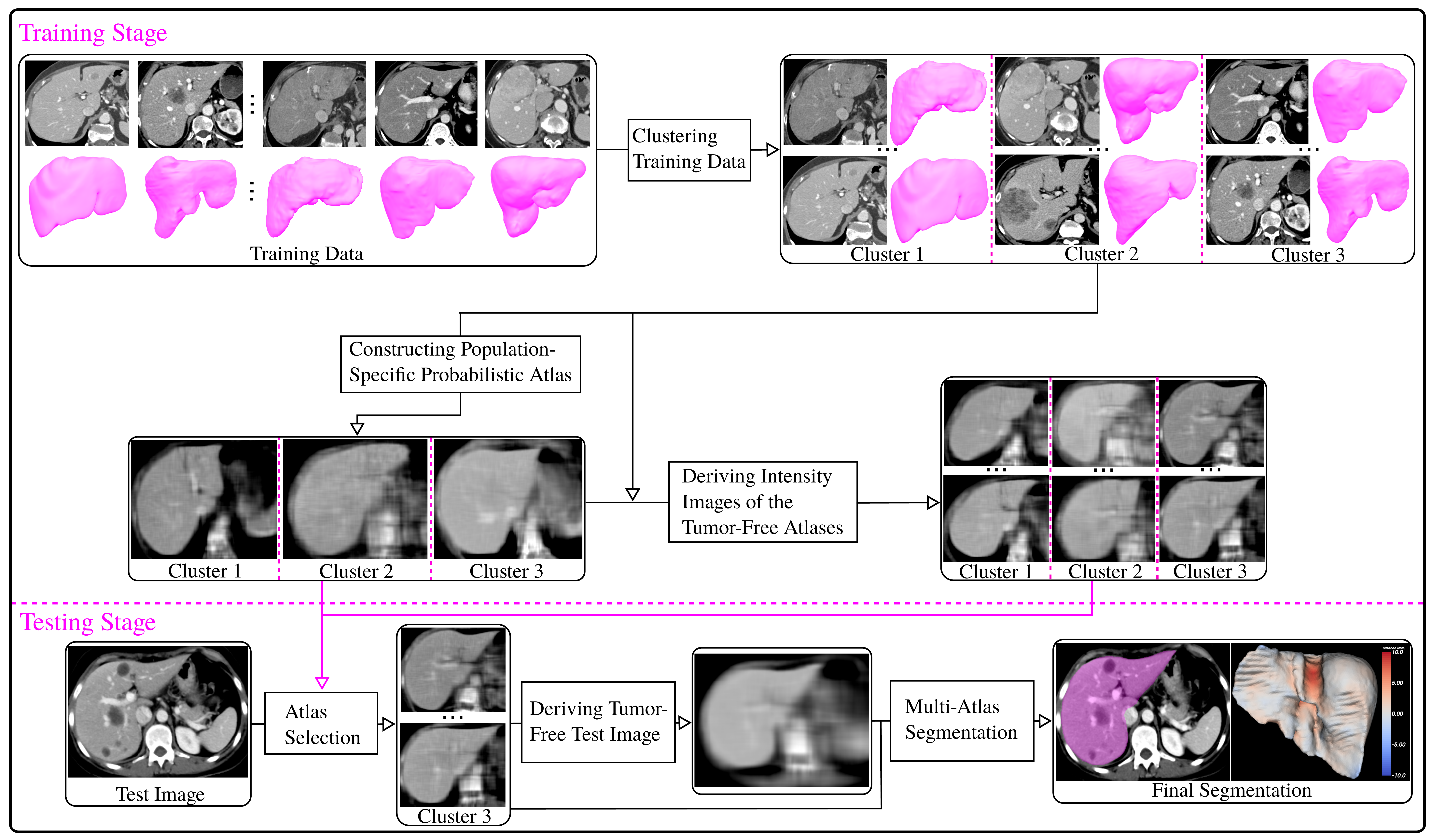}	
		\caption{The main workflow of the proposed framework 
		         for pathological liver CT segmentation, 
		         consisting of training and testing stages.}
		\label{fig:LRSCC}
	\end{center}
\end{figure*}

The main workflow of the proposed segmentation framework is depicted 
in Fig. \ref{fig:LRSCC}.
%
%
In the training stage,
two models used in the MAS framework are constructed: 
population-specific probabilistic atlas (PA) and tumor-free liver atlases
(Section \ref{subsec:LRTD-PAs}).
In the testing stage, 
for a given test abdominal CT image,
we first implement an atlas selection strategy 
to derive patient-specific liver atlases,
then generate a tumor-free test image
based on the selected tumor-free liver atlases,
and finally perform the main steps of the MAS algorithm 
to extract the liver tissue from the tumor-free test image 
(Section \ref{subsec:LRTD-MAS}).

%

\subsection{ Low-Rank Tensor Decomposition for Liver Atlas Construction  }
\label{subsec:LRTD-PAs}  


To derive patient-specific liver atlases, 
and to substantially reduce the computational cost
of the MAS framework, 
we propose an LRTD-based liver atlas construction method.
Specifically, 
we first group the training data into multiple smaller clusters corresponding
to different type cases,
then construct a population-specific PA and 
tumor-free liver atlases for each cluster separately.


\subsubsection{ Population-Specific PA and Tumor-Free Liver Atlases }
\label{subsec:LRTD-PA}



\responsess{
We first employ the same spectral clustering based algorithm~\citep{Luxburg07}
as in~\cite{shih17_MIA}
to divide the training data into multiple smaller partitions.
%
The algorithm involves two major steps:
(1) training data alignment: 
Since the affinity values used in the clustering step 
are calculated from both intensity images and liver shapes, 
the training data needs to be first aligned; 
and (2) data clustering using the spectral clustering algorithm: 
Specifically, the features used for clustering 
are the affinity values between two training samples, 
defined using harmonic mean of the intensity image and liver shape distances.
%
%
%
%
%
Please refer to~\cite{shih17_MIA} 
for a detailed description of the procedures.
}

After the training data is partitioned into multiple clusters, 
we construct a population-specific PA and 
tumor-free liver atlases for each cluster separately. 
Nevertheless, most of the training images include major pathology.
%
To mitigate the performance degradation of liver segmentation 
due to the presence of tumors in the constructed atlases,
we cast the procedure of generating tumor-free liver atlases as
an LRTD problem~\citep{Kolda11}, 
also known as TRPCA~\citep{Lu20}.
Specifically, 
we propose an LRTD-based liver atlas construction method, 
called LRTD-PA,
based on the following two empirical observations:
(1) the aligned training images are linearly correlated with each other
and form a low-rank third-order tensor~\citep{Sagheer19}; 
and (2) the portions that cannot be 
represented by the low-rank part
are the gross errors or outliers (e.g., tumors),
which can also be considered
sparse compared to the whole image tensor.
%
%
%

Let $\{I_i~|~i = 1, . . ., N_{c} \}$ be the pre-aligned training images 
with their corresponding label images $\{L_i~|~i = 1, . . ., N_{c} \}$
of cluster $c$.
We represent each training image $I_i$ as
a third-order image tensor $\bm{\mathcal{D}}_i \in \mathbb{R}^{w \times h \times d}$ 
by stacking all voxel intensity values of 
the axial slices frontal-slice-wisely, 
where $w$, $h$ and $d$ denote the width, 
height and the number of axial CT slices, respectively. 
Then we construct a third-order image repository tensor $\bm{\mathcal{X}} \in \mathbb{R}^{w \times h \times (dN_{c})}$ 
by concatenating all training image tensors
$\{\bm{\mathcal{D}}_i~|~i = 1, . . ., N_{c} \}$ frontal-slice-wisely.
Mathematically, 
the LRTD model decomposes the image repository tensor $\bm{\mathcal{X}}$
into two components according to
the following minimization: 
\begin{equation}
\label{eq:LRSD-PCPaaaa}
	(\hat{\bm{\mathcal{L}}}, \hat{\bm{\mathcal{E}}}) =  
\operatorname*{arg\,min}_{\bm{\mathcal{L}}, \bm{\mathcal{E}}} 
	                        ~\operatorname*{rank}(\bm{\mathcal{L}}) + \lambda\|\bm{\mathcal{E}}\|_0 ~~\operatorname*{s.t.}~~
\bm{\mathcal{X}} = \bm{\mathcal{L}} + \bm{\mathcal{E}},	
\end{equation}
where $\bm{\mathcal{L}}$ represents the low-rank component corresponding to the tumor-free training images via tensor rank minimization,
$\bm{\mathcal{E}}$ represents the sparse component corresponding to the sparse tumors via $\ell_0$-norm minimization,
%
%
%
and $\lambda$ is a trade-off factor
between the two components. 
Since $\bm{\mathcal{E}}$ is employed to explicitly model the sparse gross errors
via the $\ell_0$-norm,
the LRTD model fits the 
purpose of generating tumor-free training images very well,
and our proposed LRTD-PA
is robust to handle major pathology.
%
%

Nevertheless, the minimization problem in Eq. \ref{eq:LRSD-PCPaaaa} 
is known to be computationally intractable (NP-hard),
due to the non-convex property of the tensor rank and $\ell_0$-norm~\citep{Hillar13,Natarajan95}. 
Fortunately, it has been proven that 
solving the following relaxed dual convex minimization problem, 
called Tensor Principal Component Pursuit (TPCP)~\citep{Zhang20_PAMI,Lu20}, 
can achieve the same decomposition accuracy:
\begin{equation}
\label{eq:LRSD-TPCP}
	(\hat{\bm{\mathcal{L}}}, \hat{\bm{\mathcal{E}}}) =  
\operatorname*{arg\,min}_{\bm{\mathcal{L}}, \bm{\mathcal{E}}} 
	                        ~\|\bm{\mathcal{L}}\|_{*} + \lambda\|\bm{\mathcal{E}}\|_1 ~~\operatorname*{s.t.}~~
\bm{\mathcal{X}} = \bm{\mathcal{L}} + \bm{\mathcal{E}},	
\end{equation}
where the tensor nuclear norm $\|\cdot\|_{*}$
is defined as in Section \ref{subsec:tensor-preliminaries}.
The tensor nuclear norm and $\ell_1$-norm 
are the convex surrogates of 
the tensor rank and $\ell_0$-norm, respectively.
Under certain incoherence conditions, 
it has been proven that TPCP can 
exactly recover the underlying low-rank $\bm{\mathcal{L}}$ 
and sparse $\bm{\mathcal{E}}$ components  
with high probability~\citep{Lu19_TRPCA}.

However, empirically we find that
when performing the LRTD directly on the image repository tensor $\bm{\mathcal{X}}$
that consists of all slices of the training images,
the results are unsatisfactory.
It is mainly because
the differences between consecutive image slices
in terms of both background and 
liver tissue accumulate along the axial direction,
resulting in overall rapid changes over the whole image volume.
The image repository tensor $\bm{\mathcal{X}}$
thus does not possess strong low-rank property any more.
We hypothesize that performing the LRTD on 
smaller segments consisting of multiple consecutive image slices 
can lead to better results,
since smaller segments exhibit fewer overall changes,
they will lie on a low-rank subspace~\citep{Lee18}.
Furthermore, 
both the computational cost and memory usage
will be reduced substantially.
To this end, 
we propose a multi-slice LRTD scheme 
to recover the underlying low-rank structure embedded in 3D medical images.
Specially,
we first partition each training image tensor $\bm{\mathcal{D}}_i$
into smaller segments $\{\bm{\mathcal{D}}_{ij}~|~j = 1, . . ., N_{s} \}$ 
consisting of multiple consecutive image slices of length $K$, 
then construct a third-order image repository tensor $\bm{\mathcal{X}}_{j}$ 
for each segment 
$\{\bm{\mathcal{D}}_{ij} ~|~i = 1, . . ., N_{c} \}$ frontal-slice-wisely,
and finally perform the LRTD on each segment tensor $\bm{\mathcal{X}}_{j}$ sequentially.
Mathematically, the multi-slice LRTD scheme
solves the following minimization  
for each segment tensor $\bm{\mathcal{X}}_{j}$:
\begin{equation}
\label{eq:LRSD-TPCP2}
	(\hat{\bm{\mathcal{L}}_{j}}, \hat{\bm{\mathcal{E}}_{j}}) =  
\operatorname*{arg\,min}_{\bm{\mathcal{L}}_{j}, \bm{\mathcal{E}}_{j}} 
	                        ~\|\bm{\mathcal{L}}_{j}\|_{*} + \lambda\|\bm{\mathcal{E}}_{j}\|_1 ~~\operatorname*{s.t.}~~
\bm{\mathcal{X}}_{j} = \bm{\mathcal{L}}_{j} + \bm{\mathcal{E}}_{j}.	
\end{equation}

\begin{figure*}[!t]
	\begin{center}
		\includegraphics[width=1.1\textwidth]{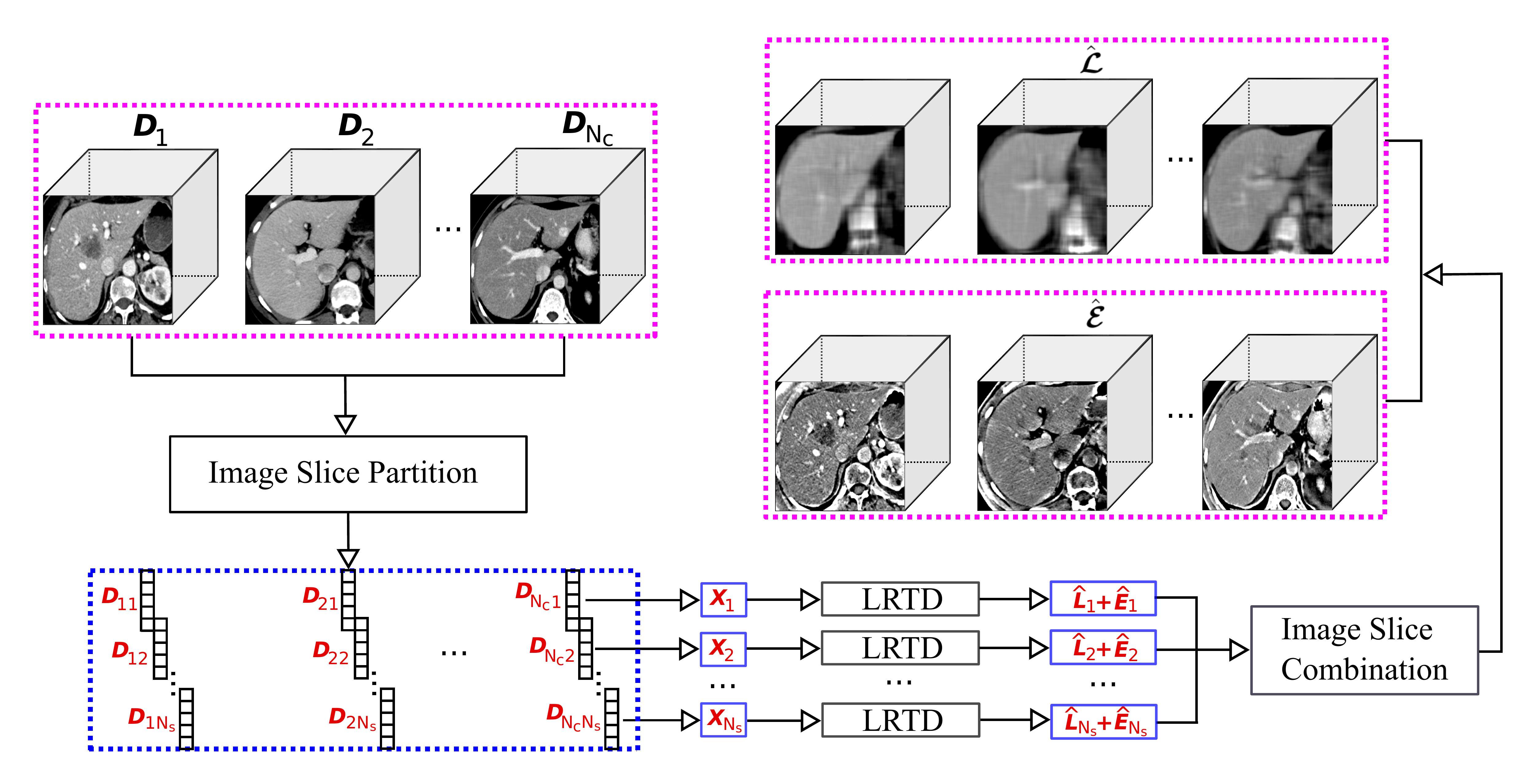}	
		\caption{An example of the multi-slice LRTD 
		applied to training images of cluster $c$.
		All training image tensors $\{\bm{\mathcal{D}}_i~|~i = 1, . . ., N_{c} \}$ 
		are first partitioned 
		into smaller segments $\{\bm{\mathcal{D}}_{ij}~|~j = 1, . . ., N_{s} \}$
		consisting of multiple consecutive image slices of length $K=5$, 
		then the LRTD is performed on each segment tensor $\bm{\mathcal{X}}_{j}$ 
		sequentially,
		and finally the decomposed low-rank and sparse components 
		of all the segment tensors
		are separately combined
        to obtain tumor-free training images $\hat{\bm{\mathcal{L}}}$ 
		and tumor images $\hat{\bm{\mathcal{E}}}$.
		}
		\label{fig:pic-segments}
	\end{center}
\end{figure*}
%
%
%
%

Eq. \ref{eq:LRSD-TPCP2} is the optimization problem of 
proposed liver atlas construction method LRTD-PA, 
through which the tumor-free training images  
lie in the low-rank component $\hat{\bm{\mathcal{L}}_{j}}$, 
while the tumors are extracted in $\hat{\bm{\mathcal{E}}_{j}}$.
After obtaining 
$\{(\hat{\bm{\mathcal{L}}_{j}}, \hat{\bm{\mathcal{E}}_{j}})~|~j = 1, . . ., N_{s} \}$,
we separately stack them frontal-slice-wisely
to obtain $\hat{\bm{\mathcal{L}}}$ and $\hat{\bm{\mathcal{E}}}$.
The tumor-free training images in the low-rank component $\hat{\bm{\mathcal{L}}}$ 
are employed to construct the population-specific PA $\bar{I}_c$,
from which we can then obtain intensity images of the tumor-free liver
atlases $\{\hat{I_i}~|~i = 1, ..., N_c\}$.
Fig. \ref{fig:pic-segments} shows 
an example of the multi-slice LRTD
applied to the training images of cluster $c$.

\responsess{
As in ~\citet{Liu15} and ~\citet{shih17_MIA},
we integrate the multi-slice LRTD into an iterative registration framework 
to construct the population-specific PAs and the tumor-free liver atlases, 
as summarized in Algorithm~\ref{algo:population-specific-PA}.
Specifically, 
during each registration iteration, 
the low-rank liver component and the sparse tumor component 
are obtained via the multi-slice LRTD. 
Then the low-rank component is used for the next registration iteration.
Compared to directly registering a liver image 
with major pathology to the template, 
much more accurate registration can be achieved 
by its low-rank liver component, 
where the pathology is reduced or eliminated. 
The decomposed low-rank and sparse components 
will be improved as more registration iterations are performed.
Therefore, 
the LRTD makes the iterative registration framework 
much more robust to the presence of major pathology in liver images, 
leading to more accurate results of atlas construction 
and atlas-based segmentation.
%
%
}
Note that 
since the LRTD is performed in a segment-by-segment fashion, 
to maintain a smooth transition between neighboring segments 
in the tumor-free training images, 
we expand each segment to overlap with one 
neighboring slice at each end, 
and average the decomposition results in overlapping slices.
Thus the smallest possible segment length $K$ is 2.



\begin{algorithm}[!t]
	\begin{center}	
    
    \begin{algorithmic}[0]
	

	\textbf{Input:}~Training images of class $c$: $\{I_i~|~i = 1, ..., N_c\}$, and the maximum number of iterations: $N_{max}$.\\
	\smallskip
    \textbf{Output:}~The probability atlas $\bar{I}_c$, and intensity images of the tumor-free liver atlases $\{\hat{I_i}~|~i = 1, ..., N_c\}$ for class $c$. \\
    
	\medskip	
	\STATE{ \textbf{1.} Select the training image containing smallest tumor region as the initial template $\bar{I}^0$. }\\		
	\smallskip

			\STATE{ \textbf{2.} Compute the non-rigid transformation $T_{i}^0$ that warps ${I_i}$ to the template $\bar{I}^0$: 			
			$I_i^1 \leftarrow T_{i}^0( {I_i} ), i = 1, ..., N_c$, and obtain the corresponding third-order image tensors:
	$\bm{\mathcal{D}}_i^1 \leftarrow I_i^1, i = 1, ..., N_c.$ }\\	
		
	\medskip	
	
    \STATE{ \textbf{3.} Obtain the population-specific probability atlas iteratively:} \\
	\FOR{$k=1$ \TO $N_{max}$}
	
		\STATE{  \hspace{0.3em}  \textbf{3.1} Compute low-rank parts via multi-slice LRTD on $\bm{\mathcal{X}}$: \\ 
		        \hspace{1.7em} $[I_{LR_1}^k, I_{LR_2}^k, ..., I_{LR_{N_c}}^k]~\leftarrow~\texttt{MS-LRTD(}~\bm{\mathcal{X}} = [\bm{\mathcal{D}}_1^k, \bm{\mathcal{D}}_2^k, ..., \bm{\mathcal{D}}_{N_c}^k ]~\texttt{)}$}.
				
		\smallskip
		\STATE{ \hspace{0.3em} \textbf{3.2} Obtain the new template using low-rank parts: $\bar{I}^k \leftarrow \frac{1}{N_c} \sum_{n=1}^{N_c} {I_{LR_i}^k}$.} \\
		\medskip
		
			\STATE  \hspace{0.3em} \textbf{3.3}  Compute the non-rigid transformation $T_{i}^k$ that warps ${I_{LR_i}^k}$ to the template $\bar{I}^k, i = 1, ..., N_c$.
			
			\smallskip
			\STATE{   \hspace{0.3em} \textbf{3.4}  Align ${I_i^k}$ to the space of the template $\bar{I}^k$ using $T_{i}^k$: $I_i^{(k+1)} \leftarrow T_{i}^k( {I_i^k} ), i = 1, ..., N_c$, and obtain the corresponding third-order image tensors: $\bm{\mathcal{D}}_i^{(k+1)} \leftarrow I_i^{(k+1)}, i = 1, ..., N_c.$} \\	
		
    \ENDFOR
	
	\smallskip
	
	$\bar{I}_c \leftarrow \bar{I}^{N_{max}}$. \\

	\medskip
	\STATE{ \textbf{4.} Generate the tumor-free liver atlases of class $c$: } \\
	
	     \STATE{ \hspace{1em} \textbf{4.1} Compute the non-rigid transformation $T_{i}$ that warps ${I_i}$ to the probability atlas $\bar{I}_c$:
		$ I_i^{'} \leftarrow T_{i}( I_i ), i = 1, ..., N_c$, and obtain the corresponding third-order image tensors: $\bm{\mathcal{D}}^{'}_i \leftarrow I^{'}_i, i = 1, ..., N_c.$}
	
	\smallskip
	\STATE{ \hspace{1.0em}  \textbf{4.2} Compute low-rank parts via multi-slice LRTD on $\bm{\mathcal{X}}^{'}$: \\ \hspace{2.5em} $[I_{LR_1}^{'}, I_{LR_2}^{'}, ..., I_{LR_{N_c}}^{'}]~\leftarrow~\texttt{MS-LRTD(}~\bm{\mathcal{X}}^{'} = [\bm{\mathcal{D}}_1^{'}, \bm{\mathcal{D}}_2^{'}, ..., \bm{\mathcal{D}}_{N_c}^{'} ]~\texttt{)}$}.\\

	\smallskip
	
		\STATE \hspace{1em}  \textbf{4.3}  $ \hat{I_i} \leftarrow T_{i}^{-1}( I_{LR_i}^{'} ), i = 1, ..., N_c. $

    \end{algorithmic}

		\caption{Population-Specific Probabilistic Atlas and 
		         Tumor-Free Liver Atlases Construction Procedure }		 
				 
        \label{algo:population-specific-PA}
	\end{center}
\end{algorithm}


\subsubsection{ Optimization }
\label{subsec:MLR-SCC}   

In the past few years, many optimization algorithms have been proposed to 
solve the LRTD problem~\citep{Sidiropoulos17},  
and stable recovery of the low-rank $\bm{\mathcal{L}}$ 
and sparse $\bm{\mathcal{E}}$ components 
can be guaranteed~\citep{Lu19_TRPCA,Zhang20}. 
To achieve both efficiency and scalability, 
we use the
alternating direction method of multipliers (ADMM) algorithm~\citep{Boyd10,Lu20}
to solve the TPCP problem in Eq. \ref{eq:LRSD-TPCP2}. 
The ADMM algorithm is
a first-order optimization method 
where both the objective function and the constraints
exhibit separable structures.
It has been widely utilized for solving convex optimization problems
with well-established convergence properties~\citep{Boyd10}.
Algorithm~\ref{algo:ADMM} (\ref{sec_Appen:closed-form-ADMM})
summarizes
the derived ADMM algorithm 
for solving the TPCP problem (Eq. \ref{eq:LRSD-TPCP2}) 
in the proposed liver atlas construction method LRTD-PA. 
The detailed mathematical derivation procedure
is given in \ref{sec_Appen:closed-form-ADMM}.
%
Algorithm~\ref{algo:ADMM2} (\ref{sec_Appen:closed-form-ADMM})
shows
the optimization procedure of the proposed multi-slice LRTD scheme.
%
%
%
The convergence properties of the ADMM algorithm 
with two blocks, as in this study,
have been well established~\citep{Boyd10}. 
Suppose that the size of the data tensor 
$\bm{\mathcal{X}}$ is $n_1 \times n_2 \times n_3$ 
with $n_1 \geq n_2$,
the main computational cost of the ADMM algorithm
lies in computing the t-SVT operator $\mathbf{D}_{\tau}$ 
in the $\bm{\mathcal{L}}$ subproblem 
(Eq. \ref{eq:LRSD-A-solution} in \ref{sec_Appen:closed-form-ADMM}).
For any general invertible matrix $\mathbf{M} \in \mathbb{R}^{n_3 \times n_3}$,
the per-iteration computational complexity of the ADMM algorithm is 
$O(n_1n_2{n_3}^2 + n_1{n_2}^2{n_3})$~\citep{Lu19_TRPCA}.

\subsection{ LRTD-Based MAS Algorithm }
\label{subsec:LRTD-MAS}  

After the tumor-free liver atlases are constructed, 
we propose an LRTD-based MAS algorithm to
perform liver CT segmentation.
Specifically, given a test image $I_t$,
we first implement an atlas selection strategy
to obtain patient-specific liver atlases,
and to substantially improve the computational efficiency.
Then based on intensity images of 
the selected tumor-free liver atlases, 
we generate a tumor-free test image $I_{LR_t}$.
Finally,
the main steps of the MAS algorithm are performed  
to extract the liver tissue from $I_{LR_t}$.

Considering that the liver tissue exhibits high anatomical variability, 
we construct patient-specific liver atlases for each test image 
via an atlas selection strategy~\citep{shih17_MIA}
to achieve more accurate liver segmentation. 
%
%
%
\responsess{
Specifically, 
we first find the most suitable cluster in the training dataset 
according to the normalized cross correlation ($NCC$) similarity measure given by:
\begin{equation}
\label{eq:MLR-SSC}
	NCC( I_i , I_j  ) = 
	\frac{ \sum_{ \mathbf{x} \in \Omega } { \left( I_i(\mathbf{x}) - \bar{I_i}\right) \left(I_j(\mathbf{x}) - \bar{I_j}\right)  } }
		 { \sqrt{ \sum_{ \mathbf{x} \in \Omega } { \left(I_i(\mathbf{x}) - \bar{I_i}\right)^2 } } \sqrt{ \sum_{  \mathbf{x} \in \Omega  } { \left(I_j(\mathbf{x}) - \bar{I_j}\right)^2  } }  } , 
\end{equation}
%
where $\bar{I_i}$ and $\bar{I_j}$ 
denote the mean values of the two images $I_i$ and $I_j$
within the overlapping region $\Omega$. 
The training data cluster $c$ is selected 
when the $NCC$ between 
$I_t$ and
the $\bar{I}_c^{'}$ of the population-specific PA 
after its warping into the space of $I_t$,
i.e., $NCC (\bar{I}_c^{'}, I_t)$, is the largest.
}
%
%
Then the tumor-free intensity images and their corresponding label images
within the chosen cluster $c$
are regarded as the patient-specific liver atlases 
for the test image $I_t$.

\begin{figure*}[!t]
	\begin{center}
		\includegraphics[width=1.0\textwidth]{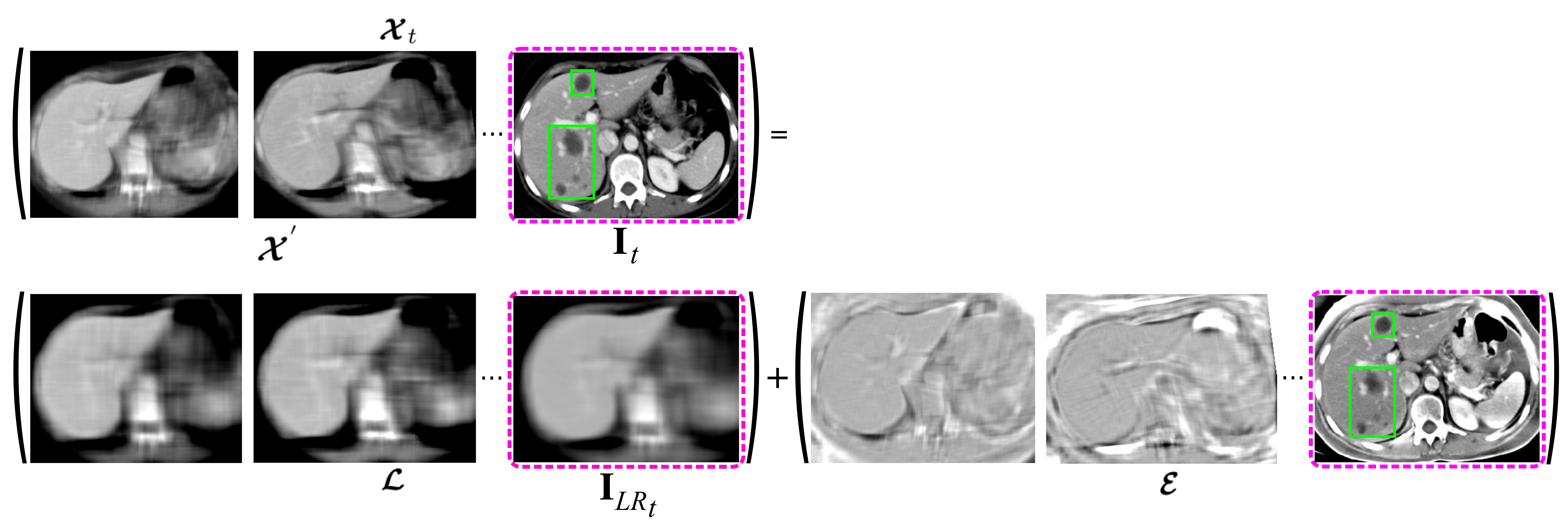}	
		\caption{An example of generating the tumor-free image
				 $I_{LR_t}$ for a test liver image $I_t$
				 (indicated by the pink dashed-line rectangle in $\bm{\mathcal{X}}_t$) 
		         using the proposed low-rank tensor decomposition based
 				 probabilistic atlas (LRTD-PA).
				 The data tensor $\bm{\mathcal{X}}_t$
				 is decomposed into a low-rank component $\bm{\mathcal{L}}$
 				 corresponding to the tumor-free liver images, 
				 and a sparse component $\bm{\mathcal{E}}$ corresponding to 
				 the sparse tumors
				 (indicated by green rectangles).
				 For the purpose of illustration, 
				 it only shows one specific 2D slice of the CT volume.}
		\label{fig:LRTD-PA2}
	\end{center}
\end{figure*}

Generally, the test image $I_t$ is directly employed to perform 
image registrations in the MAS algorithm.
However, we empirically find that most of the test images contain tumors.
In order to achieve accurate
pairwise image registration and label propagation,
we use the proposed LRTD-PA to generate a tumor-free test image $I_{LR_t}$.
Specifically,
given the test image $I_t$,
we first construct a third-order data tensor $\bm{\mathcal{X}}_t \in \mathbb{R}^{w \times h \times \left(d(N_{c}+1) \right)}$
by concatenating 
the warped image repository tensor of the chosen cluster $\bm{\mathcal{X}}^{'} \in \mathbb{R}^{w \times h \times (dN_{c})}$
with the test image tensor
$\bm{\mathcal{D}}_t \in \mathbb{R}^{w \times h \times d}$
frontal-slice-wisely.
After performing the multi-slice LRTD 
on the data tensor $\bm{\mathcal{X}}_t$,
we can obtain the tumor-free test image $I_{LR_t}$,
as shown in Algorithm~\ref{algo:targetImage_LRSD-PA}.
%
%
Fig. \ref{fig:LRTD-PA2} shows 
an example of generating the tumor-free image $I_{LR_t}$
for a test image $I_t$
using the proposed LRTD-PA.



\begin{algorithm}[!t]
	\begin{center}	
    
    \begin{algorithmic}[0]
	

	\textbf{Input:}~Intensity images of the tumor-free atlases within the chosen cluster $c$: $\{\hat{I_i}~|~i = 1, ..., N_c\}$, the test image: $I_t$, and the maximum number of iterations: $N_{max}$. \\
	
	\medskip
    \textbf{Output:}~The tumor-free test image $I_{LR_t}$.	\\
	
	\medskip	

			\STATE{ \textbf{1.} Compute the non-rigid transformation $T_{i}^0$ that warps $\hat{I_i}$ to the target image $I_t$: 
			%
			$\hat{I_i}^1 \leftarrow T_{i}^0( {\hat{I_i}} ), i = 1, ..., N_c$, and obtain the corresponding third-order image tensors: $\hat{\bm{\mathcal{D}}}_i^1 \leftarrow \hat{I_i}^1, i = 1, ..., N_c.$ }
			
		
	\medskip	
	
	\STATE{ \textbf{2.} Obtain the test image tensor: $\bm{\mathcal{D}}_t \leftarrow I_t.$ }
		\medskip

	\STATE{ \textbf{3.} Obtain the tumor-free test image iteratively:} \\
	\FOR{$k=1$ \TO $N_{max}$}

		\STATE{ \textbf{3.1} Compute the low-rank test image via multi-slice LRTD: \\
	    ${I_{LR_t}^k}~\leftarrow~\texttt{MS-LRTD(}~\bm{\mathcal{X}}_t = [\hat{\bm{\mathcal{D}}_1}^k, \hat{\bm{\mathcal{D}}_2}^k, ..., \hat{\bm{\mathcal{D}}_{N_c}}^k, \bm{\mathcal{D}}_t]~\texttt{)}$.}\\		
				
		
		
		\medskip
		
			\STATE{ \textbf{3.2} Compute the non-rigid transformation $T_{i}^k$ that warps $\hat{I_i}$ to $I_{LR_t}^k$: \\ 
			%
		    $\hat{I_i}^{(k+1)} \leftarrow T_{i}^k( {\hat{I_i}} ), i = 1, ..., N_c$, and obtain the corresponding third-order image tensors: $\hat{\bm{\mathcal{D}}}_i^{(k+1)} \leftarrow \hat{I_i}^{(k+1)}, i = 1, ..., N_c.$ }				
		

    \ENDFOR

	$I_{LR_t}  \leftarrow I_{LR_t}^{N_{max}}$.

    \end{algorithmic}

		\caption{Tumor-Free Test Image Derivation Procedure}
        \label{algo:targetImage_LRSD-PA}
	\end{center}
\end{algorithm}


%

After obtaining the tumor-free test image $I_{LR_t}$, 
it is the input to the MAS algorithm.
The main steps of the MAS algorithm 
(i.e., image registration, label propagation, and label fusion)
are then performed   
to obtain the liver segmentation result.

Firstly, all intensity images of the tumor-free liver atlases 
within the chosen cluster $c$
$\{\hat{I_i}~|~i = 1, ..., N_c \}$
are non-rigidly warped to the space of $I_{LR_t}$,
resulting in $N_c$ non-rigid transformations $\{T_i~|~i = 1, ..., N_c\}$.
These non-rigid transformations then 
propagate the corresponding label images 
$\{L_i~|~i = 1, ..., N_c \}$ into the space of $I_{LR_t}$,
resulting in the propagated atlas labels $\{L_i^{'}~|~i = 1, ..., N_c\}$.

Finally, the $N_c$ propagated atlas labels are combined
via label fusion
to obtain the liver segmentation result.
Specifically, we adopt the joint label fusion (JLF) algorithm~\citep{Wang13}
considering its high performance.
The JLF is a statistical local weighted label fusion algorithm, 
where the optimal weights for label fusion are obtained  
by minimizing the total expectation of labeling errors.
Moreover, the pairwise correlation between atlases is explicitly modeled 
as the joint probability of two atlases creating a segmentation error at a voxel.
Please refer to~\cite{Wang13} for more details on the JLF algorithm.
To obtain the final segmentation result from the
posterior probability map of liver likelihood produced by the JLF algorithm,
it performs a series of simple post-processing operations on the probability map,
consisting of thresholding 
using Otsu's method~\citep{Otsu79}, 
morphological opening operator to remove small unconnected components 
and the noise, 
and morphological closing operator to fill small cavities. 

%% file: chapters/results.tex

\section{ Experiments }
\label{section:Experiments}

\subsection{ Datasets }

In order to evaluate the performance of the proposed MAS framework, 
and to show its clinical applicability,
we tested it on a clinical dataset of \responsess{110} abdominal CT scans of
pathological liver cases 
from three publicly available databases.
Table \ref{table:liver_datasets} outlines
the specifications of the CT scans from the three databases.
\begin{table}[!t]
    \renewcommand{\arraystretch}{1.3}
	\caption{ \responsess{Specifications of the CT scans from the three databases.} }
    \label{table:liver_datasets}
    \centering
	
\begin{tabular}{ p{8em} p{3.5em} p{4em} p{4.5em} p{3.5em} p{4em} }
\hline
Database & Number of scans & In-plane matrix size & In-plane resolution [mm] & Number of slices  & Slice thickness [mm] \\
\hline
SLIVER07-Train  &  20  &  512$\times$512  &  0.58-0.81  &  64-394  &  0.7-5.0 \\
3Dircadb1   &  20  &  512$\times$512  &  0.56-0.86  &  74-260  &  1.0-4.0 \\
\responsess{LiTS2017-Test} &  \responsess{70}  &  \responsess{512$\times$512}  &  \responsess{0.65-1.37}  &  \responsess{75-841}  &  \responsess{0.8-5.0} \\
\hline
\end{tabular}

\end{table}

Both SLIVER07-Train~\footnote{\url{http://www.sliver07.org}}
and 3Dircadb1~\footnote{\url{http://www.ircad.fr/softwares/3Dircadb/3Dircadb1/index.php?lng=en}}
include 20 abdominal CT scans with corresponding ground truths, 
provided by the organizers of the Segmentation of the Liver
Competition 2007 (SLIVER07)~\citep{Heimann09},
and IRCAD~\citep{Soler09}, the French Research Institute against Digestive Cancer, respectively.
\responsess{The LiTS2017-Test~\footnote{\url{https://competitions.codalab.org/competitions/17094\#participate}}
consists of 70 abdominal CT scans without corresponding ground truths~\citep{LiTS17}.}
Most of the cases in the three databases were  
with hypodense and/or hyperdense tumor contrast levels, 
mainly including hepatocellular carcinoma (HCC),
metastases, hemangiomas, and cysts of different size.
The CT scans in the three databases
were acquired by a wide variety of CT scanners from
different vendors. 
%
In this study, the SLIVER07-Train database  
was employed to train the models, 
and to determine the parameter settings
of the proposed method.
The other two databases were only used 
to evaluate the segmentation performance, 
and to compare our method with state-of-the-art methods.

\subsection{ Evaluation Metrics and Statistical Hypothesis Testing }
\label{subsec:Metrics}

\responsess{
In order to perform quantitative evaluations of segmentation accuracy, 
the following volume and surface based metrics were used:
Dice coefficient ($DC$)~\citep{Dice45},
Jaccard index ($JI$)~\citep{Jaccard1901},
and average symmetric surface distance ($ASD$)~\citep{Heimann09}.}
\responsess{
The units of $DC$ and $JI$ are percent,
and the unit of $ASD$ is millimeter.
For $JI$ and $DC$ ($ASD$),
the larger (smaller) the value is, 
the more accurate the segmentation result will be.} 
 
In order to determine whether the differences in segmentation accuracy
between our method and other compared methods
were statistically significant in the experiments,
the paired $t$-test was carried out 
with a significance level of $p<0.05$.
The null hypothesis is that 
the mean values of the same evaluation metric 
are exactly the same for the compared methods.

\subsection{ Implementation Details }
\label{subsec:Metrics}

To reduce image noise
without deteriorating the
important edge information,
it first preprocessed 
all training and test images
via the 3D anisotropic diffusion filter~\citep{PM90}.
To perform the pairwise non-rigid image registrations,
it utilized the publicly available elastix toolbox~\footnote{\url{http://elastix.isi.uu.nl}}~\citep{Klein10},
in which the cubic B-spline based free-form deformation (FFD) model~\citep{Rueckert99} with
normalized mutual information ($NMI$)~\citep{Studholme99} 
as the similarity measure
was applied.
%
%
%
The JLF algorithm implemented in the open source 
PICSL Multi-Atlas Segmentation Tool
~\footnote{\url{https://www.nitrc.org/projects/picsl_malf}}~\citep{Wang132}
was employed to perform label fusion. 
%
%
In Algorithm~\ref{algo:population-specific-PA}, 
the masked training images
via liver binary masks
were used to obtain all the non-rigid transformations,
which were then propagated to the original training images. 
While only the original training and test images were used 
in Algorithm~\ref{algo:targetImage_LRSD-PA}.
%
%

The parameter settings for 
the proposed MAS framework
were optimized
via leave-one-out cross-validation (LOOCV)
using the SLIVER07-Train database.
%
Table \ref{table:Parameters} lists
the parameter settings for the proposed segmentation framework.

\begin{table}[!t]
    \renewcommand{\arraystretch}{1.3}
    \caption{ Parameter settings for the proposed segmentation framework. }
    \label{table:Parameters}
    \centering
	
\begin{tabular}{ p{4.5em} p{2.2em} p{21em} }
\hline
Parameter & Value & Description  \\
\hline
$k$ &  3   &  Number of training data clusters for LRTD-PA \\ 
 $K$ &  5   &  Number of consecutive image slices for the multi-slice LRTD scheme\\ 
 $\lambda$ &  $\lambda_0$   &  Parameter in Eq. \ref{eq:LRSD-TPCP2} for LRTD-PA. $\lambda_0 = 1 /\sqrt{ \operatorname*{max}(n_1, n_2)n_3}$ is the default value for $\lambda$ as suggested in~\citet{Lu20}, 
where $(n_1, n_2, n_3)$ is the size of the data tensor.\\
 $N_{max}$ &  3    &  Maximum number of iterations in Algorithm~\ref{algo:population-specific-PA} and Algorithm~\ref{algo:targetImage_LRSD-PA}   \\
\hline

\end{tabular}

\end{table}

In the experiments, it compared the proposed LRTD-based MAS framework 
with two other closely related MAS frameworks: 
conventional MAS and the LRMD-based MAS.
The latter one can be considered as a special case of the proposed framework,
by substituting LRTD with LRMD to generate tumor-free liver images. 
Refer to~\citet{shih17_MIA} for more details.
%
%
%
%
To make the comparisons between different frameworks fair, 
the same training data clustering,
non-rigid registration, 
and the JLF steps were utilized.
Furthermore, we compared the proposed segmentation framework with
other state-of-the-art methods.

All the MAS frameworks were implemented
in Python on Ubuntu 18.04.
The SimpleITK library~\citep{Yaniv18}
was utilized to perform the image processing.
\responsess{All the tests in this study were run on a PC 
equipped with an Intel Core i7 processor, 
32 GB RAM, 
and an NVIDIA GTX 1650 GPU.}
The average time to segment one test image
was about \SI{95}{\minute},
most of which was spent 
performing the pairwise non-rigid image registrations and the JLF,
with an average time of about \SI{3}{\minute} (per image registration) 
and \SI{15}{\minute}, respectively.
%

\section{ Results }
\label{section:Results}

\subsection{ Parameter Settings for the Liver Atlas Construction Method LRTD-PA }

\subsubsection{ The Transform Matrix $\mathbf{M}$ }
\label{subsubsection:FFT-DCT}

An important implementation in the proposed
liver atlas construction method LRTD-PA 
is to choose the best transform matrix 
$\mathbf{M}$ of $\star_{M}$-product 
in Section \ref{subsec:tensor-preliminaries}, 
which converts the data tensor $\bm{\mathcal{X}}$ 
into other transform domain as $\bar{\bm{\mathcal{X}}}$. 
%
To this end, we tested the effect of
three commonly used transforms~\citep{Kernfeld15}, 
i.e., FFT, DCT, and Daubechies-4 discrete wavelet transform (DWT),
on the image intensity standard deviation $\sigma$ 
and entropy $H$ of 
the masked tumor-free liver images via liver binary masks.
Since the smaller the $\sigma$ and $H$ are,
the more homogeneous and smooth the appearance 
of the generated tumor-free liver image will be~\citep{Liu15}.

Table \ref{table:transform_matrix} shows the $\sigma$ and $H$
of the masked tumor-free liver images
with different choices of the transform matrix $\mathbf{M}$
using the SLIVER07-Train database.
Although all the three transform matrices
largely reduce the values of $\sigma$ and $H$  
compared to their initial values of $32.44$ and $1.38$,
respectively,
the DCT achieves the smallest mean $\sigma$ of $15.38$,
which is less than half of its initial value.
Also, 
the DCT yields the smallest mean $H$ of $1.13$.
The time needed to
perform the multi-slice LRTD on each image segment
is also given in Table \ref{table:transform_matrix}. 
The DCT is the most efficient and only takes \SI{63.68}{\milli\second}, 
while FFT costs nearly three times of that for the same task.
Therefore,
we choose the DCT as the transform matrix $\mathbf{M}$
for the proposed LRTD-PA.
%

\begin{table}[!t]
    \renewcommand{\arraystretch}{1.3}
    \caption{ Quantitative comparative results of the masked tumor-free liver images with different choices of the transform matrix $\mathbf{M}$ using the SLIVER07-Train database.			  
			}
    \label{table:transform_matrix}
    \centering
	
\begin{tabular}{ p{5.5em} p{5.5em} p{5.5em} p{5.5em} p{5.5em}}
\hline
Metrics       & Initial & FFT & DWT & DCT\\
\hline
$\sigma$     &  32.44$\pm$8.99 &  17.02$\pm$4.51 &  16.64$\pm$4.39 & \textbf{15.38$\pm$4.17}   \\
$H$ [bits]     &  1.38$\pm$0.37  &  1.23$\pm$0.32 &  1.20$\pm$0.31 & \textbf{1.13$\pm$0.30} \\
Time [\SI{}{\milli\second}]    &  &  172.25$\pm$14.61 &  76.23$\pm$7.74 & \textbf{63.68$\pm$3.88}\\
\hline
\end{tabular}

\begin{tabular}{ p{31em} }
{ \footnotesize For each metric, the mean and standard
                deviation of the overall datasets are given. 
              Bold values are the best result in that column. 
}
\end{tabular}

\end{table}

\begin{figure*}[!t]
	\begin{center}
		\includegraphics[width=0.8\textwidth]{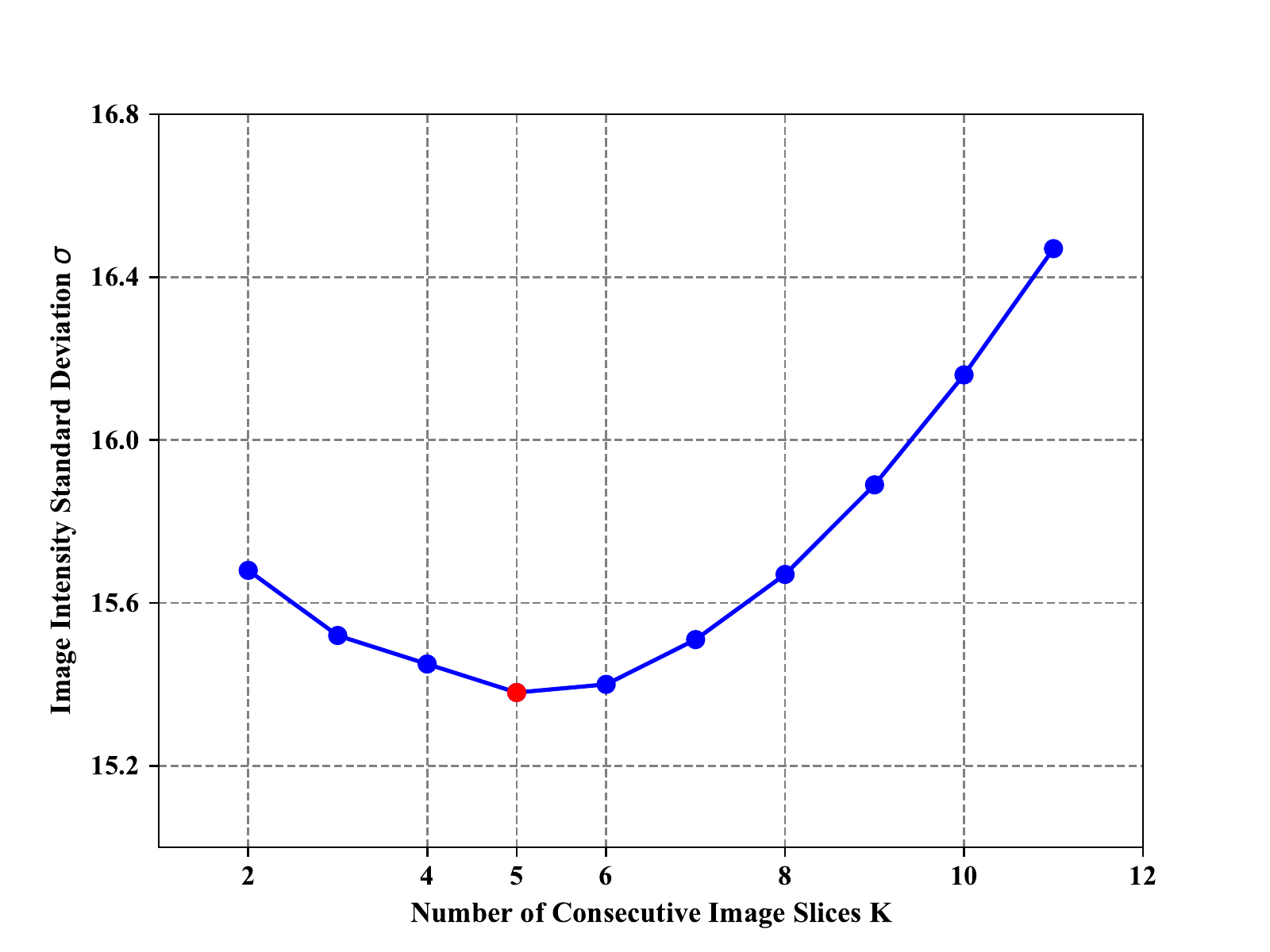}	
		\caption{Image intensity standard deviation $\sigma$ of the masked tumor-free liver images with different choices of the number of consecutive image slices $K$ for the multi-slice LRTD scheme using the SLIVER07-Train database.}
		\label{fig:K}
	\end{center}
\end{figure*}
%

\subsubsection{ The Number of Consecutive Image Slices for the Multi-Slice LRTD }
\label{subsubsection:consecutive-image-slices}

The main aim of the proposed multi-slice LRTD scheme is
to generate accurate tumor-free liver images,
while coping with the rapid changes of both background and 
liver tissue over the whole image volume.
An important hyperparameter for the multi-slice LRTD scheme 
is the number of consecutive image slices $K$,
which determines how many neighboring slices are used
to perform the LRTD.
To select the best $K$ for this study,
it tested the effect of
different choices of $K$ (from $2$ to $11$)  
on the image intensity standard deviation $\sigma$ of 
the masked tumor-free liver images via liver binary masks.

Fig. \ref{fig:K} shows the $\sigma$
of the masked tumor-free liver images
with different choices of $K$
using the SLIVER07-Train database.
It can be seen that
the value of $\sigma$ first decreases   
as the number of $K$ increases.
However, 
the value of $\sigma$ begins to increase when $K>5$, 
and the multi-slice LRTD scheme achieves 
the smallest value of $\sigma$ when $K=5$.
Therefore, 
we choose $K=5$ 
for the multi-slice LRTD scheme in this study.
Note here that
we only give the results of $\sigma$ 
with regards to different $K$ values,
since the use of image intensity entropy $H$ 
yields the same best choice $K=5$.

\subsection{ Pathological Liver Segmentation of CT Images }

In this section, to show the effectiveness of 
the proposed LRTD-based MAS framework,
we applied it to the challenging task of 
pathological liver segmentation of CT images
in the 3Dircadb1 database.
It also compared the proposed method 
with two other closely related MAS frameworks,
i.e., conventional MAS and the LRMD-based MAS,
to verify its superiority.
To make the comparisons between different MAS frameworks fair, 
the same training data clustering,
non-rigid registration, 
and the JLF steps were utilized.


\begin{figure*}[!t]
	{
		\begin{center}
			\subfloat[]
			{
			    \begin{minipage}{0.24\textwidth}
				    \includegraphics[width=1.0\textwidth]{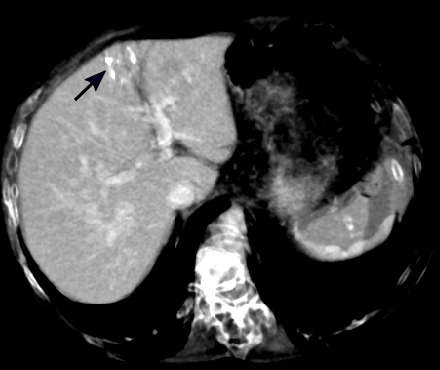}
				    \label{fig:qhull}
					
				    \includegraphics[width=1.0\textwidth]{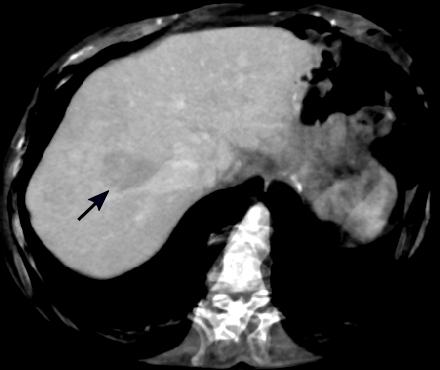}
				    \label{fig:qhull}
					
					\includegraphics[width=1.0\textwidth]{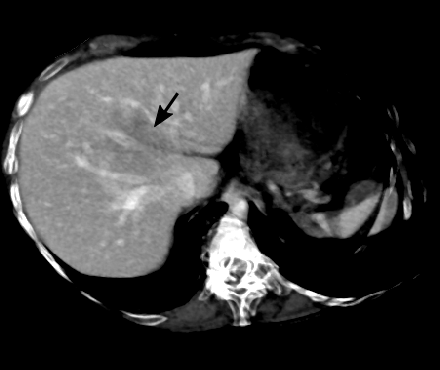}
				    \label{fig:qhull}
				\end{minipage}
			}
			\subfloat[]
			{
			    \begin{minipage}{0.24\textwidth}
				    \includegraphics[width=1.0\textwidth]{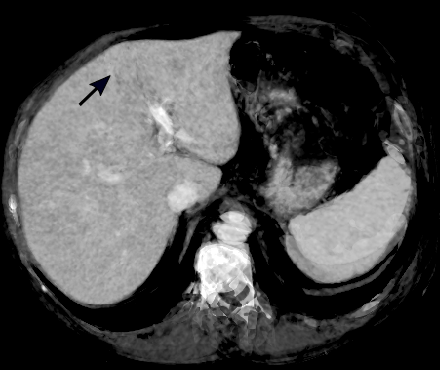}
				    \label{fig:qhull}
					
				    \includegraphics[width=1.0\textwidth]{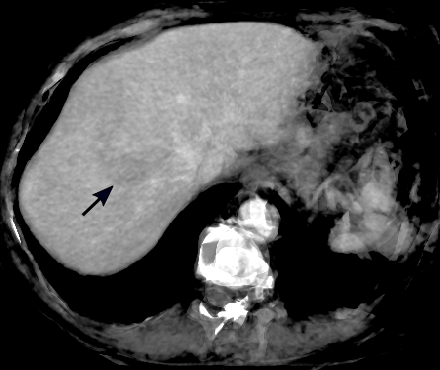}
				    \label{fig:qhull}
					
					\includegraphics[width=1.0\textwidth]{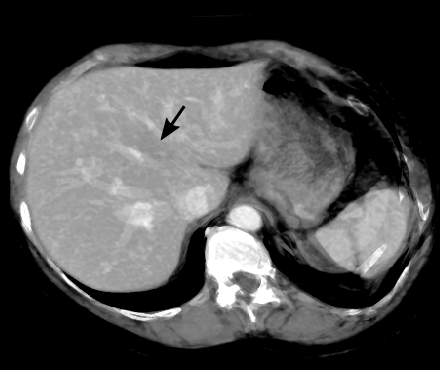}
				    \label{fig:qhull}
				\end{minipage}
			}
            \subfloat[]
			{
			    \begin{minipage}{0.24\textwidth}
				    \includegraphics[width=1.0\textwidth]{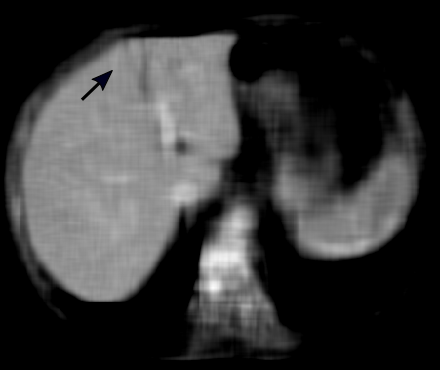}
				    \label{fig:qhull}
					
				    \includegraphics[width=1.0\textwidth]{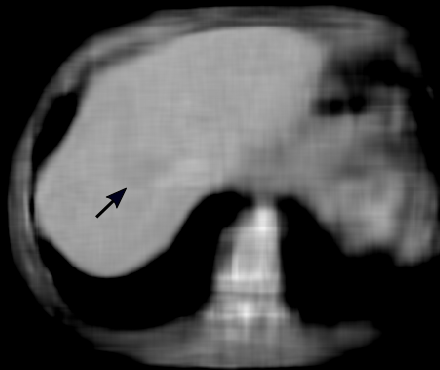}
				    \label{fig:qhull}
					
					\includegraphics[width=1.0\textwidth]{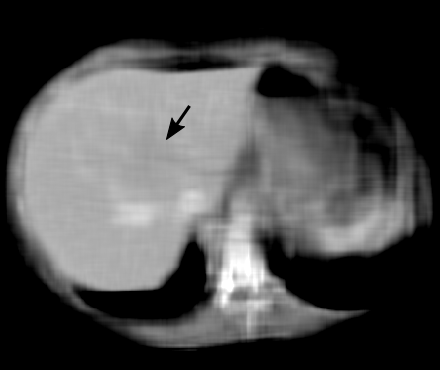}
				    \label{fig:qhull}
				\end{minipage}
			}

		\end{center}
	}
		\caption{Comparative results of the three population-specific probabilistic atlases (PAs) 	
			  generated by the conventional PA (1st column), 
			  the LRMD-PA (2nd column), 
			  and the proposed LRTD-PA (3rd column).
			  Each row shows the PA of one training data cluster.
			  The areas indicated by black arrows
			  in the PAs of the conventional PA 
			  appear totally different from that of the normal liver tissue.		  
			  }
		\label{fig:Cluster-PAss}
\end{figure*}
%

\subsubsection{ Qualitative Results }

Fig. \ref{fig:Cluster-PAss} shows 
the visually comparative results of 
the three population-specific liver PAs 
generated by the conventional PA, the LRMD-PA,
and the proposed LRTD-PA.
We can easily see that 
large areas of the PAs constructed by the conventional method  
appear totally different   
from that of the normal liver tissue 
(indicated by black arrows).
Thus, 
the conventional method 
is strongly influenced by the presence of
major pathology in training data.
In comparison, the liver tissue of the PAs constructed by 
the LRMD-PA and our method appears more homogeneous,
due to the use of  tumor-free training images,
which also leads to much more accurate pairwise image registrations.
Nevertheless, compared with the LRMD-PA,
our results are even more homogeneous and smooth,
especially in areas containing major pathology and vessels.
It is mainly because our tensor-based method 
can fully exploit 
the high correlations between neighboring slices 
of the 3D CT scans.
While the LRMD-PA   
reformats the data tensor to a matrix
by vectorizing voxel intensity values of each CT scan 
to form the column vectors,
where the local spatial information is totally lost,
and the multi-dimensional structure embedded in the tensor data is ignored,
causing severe performance degradation.
%
Therefore, our method can 
handle major pathology more effectively,
and can mitigate the performance degradation of liver segmentation 
caused by the presence of tumors in the constructed atlases.

\begin{figure*}[!t]
	{
		\begin{center}
			\subfloat[]
			{
			    \begin{minipage}{0.24\textwidth}
				    \includegraphics[width=1.0\textwidth]{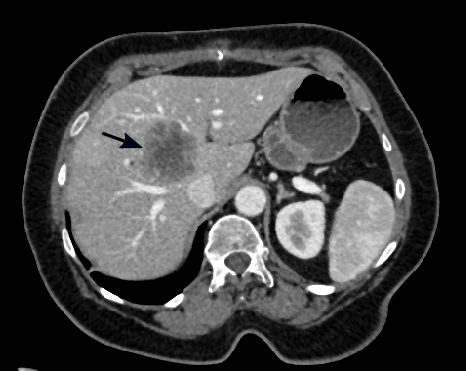}
				    \label{fig:qhull}
					
				    \includegraphics[width=1.0\textwidth]{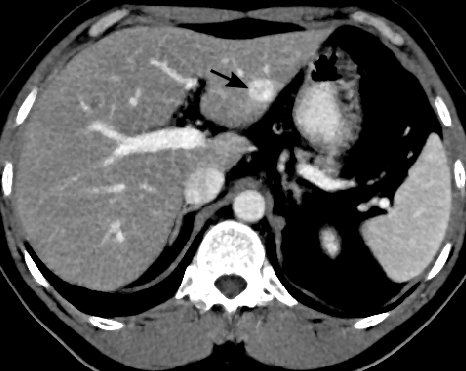}
				    \label{fig:qhull}
					
					\includegraphics[width=1.0\textwidth]{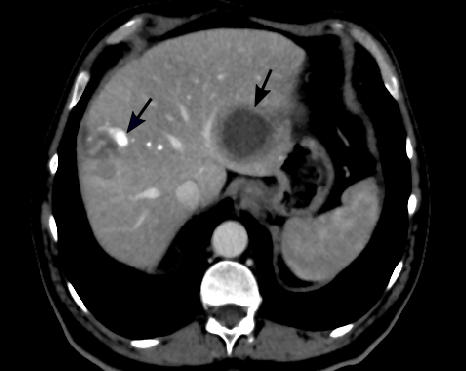}
				    \label{fig:qhull}
				\end{minipage}
			}
			\subfloat[]
			{
			    \begin{minipage}{0.24\textwidth}
				    \includegraphics[width=1.0\textwidth]{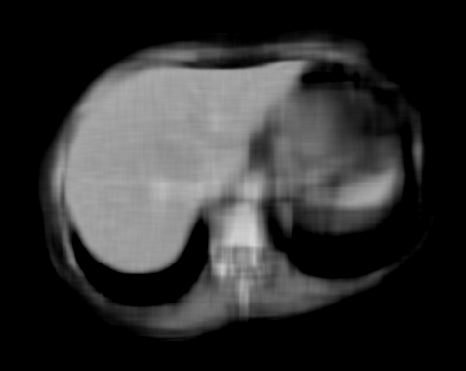}
				    \label{fig:qhull}
					
				    \includegraphics[width=1.0\textwidth]{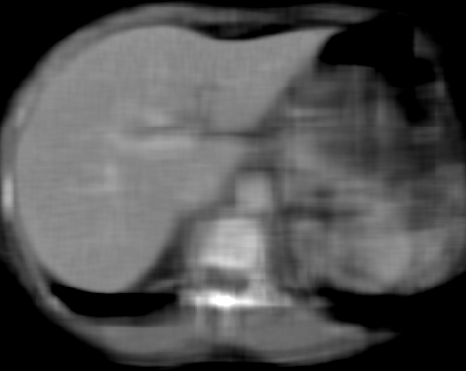}
				    \label{fig:qhull}
					
					\includegraphics[width=1.0\textwidth]{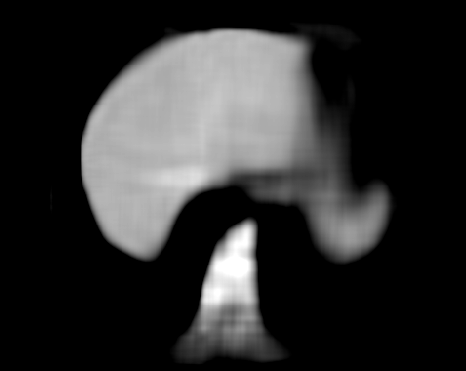}
				    \label{fig:qhull}
				\end{minipage}
			}
            \subfloat[]
			{
			    \begin{minipage}{0.24\textwidth}
				    \includegraphics[width=1.0\textwidth]{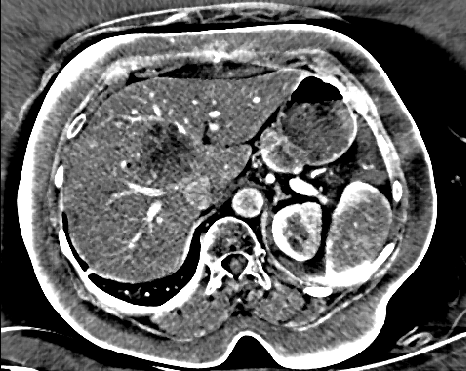}
				    \label{fig:qhull}
					
				    \includegraphics[width=1.0\textwidth]{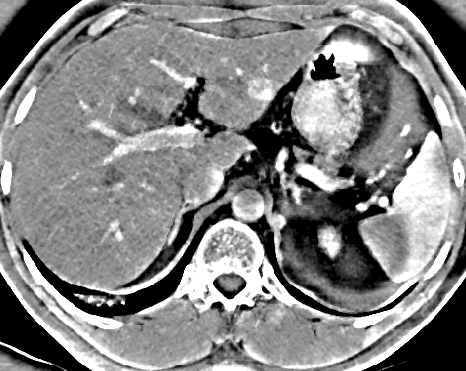}
				    \label{fig:qhull}
					
					\includegraphics[width=1.0\textwidth]{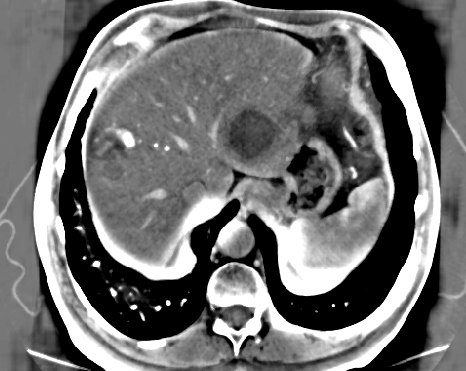}
				    \label{fig:qhull}
				\end{minipage}
			}

		\end{center}
	}	
		\caption{Results of the tumor-free liver images
			  for three challenging pathological cases 
			  generated by the proposed LRTD-PA method, 
			  consisting of liver tissue with hypodense tumor (1st row), 
			  hyperdense tumor (2nd row), 
			  and hypo- and hyperdense tumors (3rd row) 
			  indicated by black arrows.
			  In each row, (a) the original images, (b) the corresponding
			  tumor-free images (low-rank components),
			  and (c) sparse tumors (sparse components)
			  are displayed sequentially.		  
			  }
		\label{fig:tumor-correction}
\end{figure*}

Fig. \ref{fig:tumor-correction} gives 
the tumor-free liver images
for three challenging pathological cases generated by
the proposed LRTD-PA method, 
consisting of liver tissue with hypodense tumor, hyperdense tumor, 
and hypo- and hyperdense tumors
(indicated by black arrows).
%
For all the three challenging cases,
our method successfully eliminates 
the tumors with different contrast levels 
from the original images as the sparse components.
And the liver tissue in the decomposed low-rank components
appears much more homogeneous and smooth
than that in the original images.
Thus, 
by eliminating the influence of major pathology,
the proposed LRTD-based MAS algorithm
can achieve accurate pairwise image
registration and label propagation.
%
Furthermore,
it shows that
the proposed multi-slice LRTD scheme is able to successfully
recover the underlying low-rank structure embedded in 3D medical images.

%

\begin{figure*}[!t]
	{
		\begin{center}
			\subfloat[]
			{
			    \begin{minipage}{0.24\textwidth}
				    \includegraphics[width=1.0\textwidth]{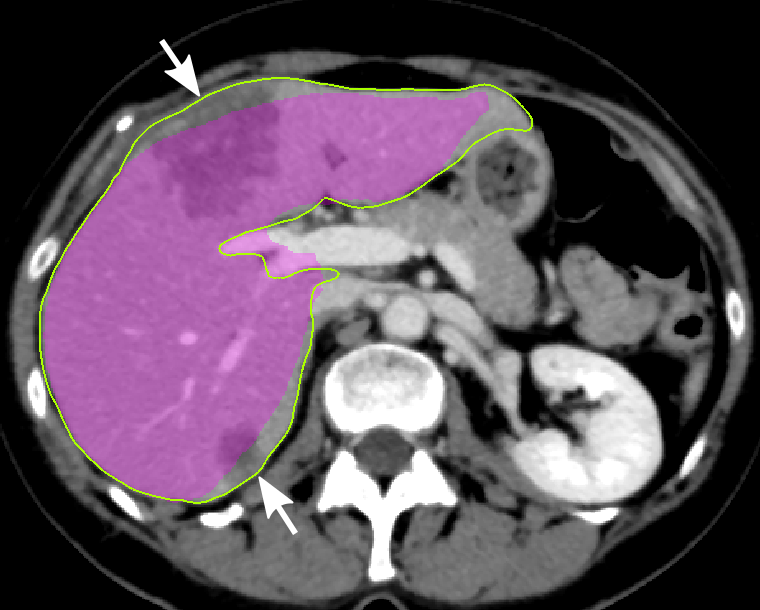}
				    \label{fig:qhull}
					
				    \includegraphics[width=1.0\textwidth]{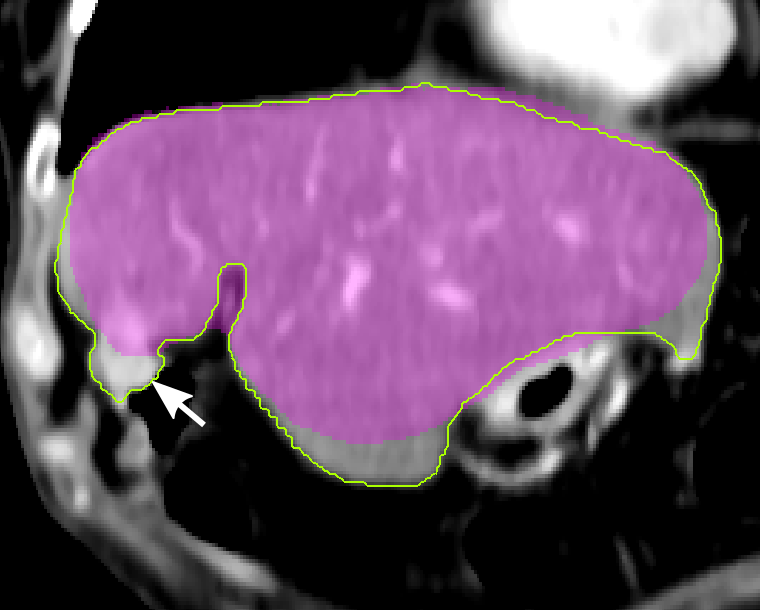}
				    \label{fig:qhull}
					
					\includegraphics[width=1.0\textwidth]{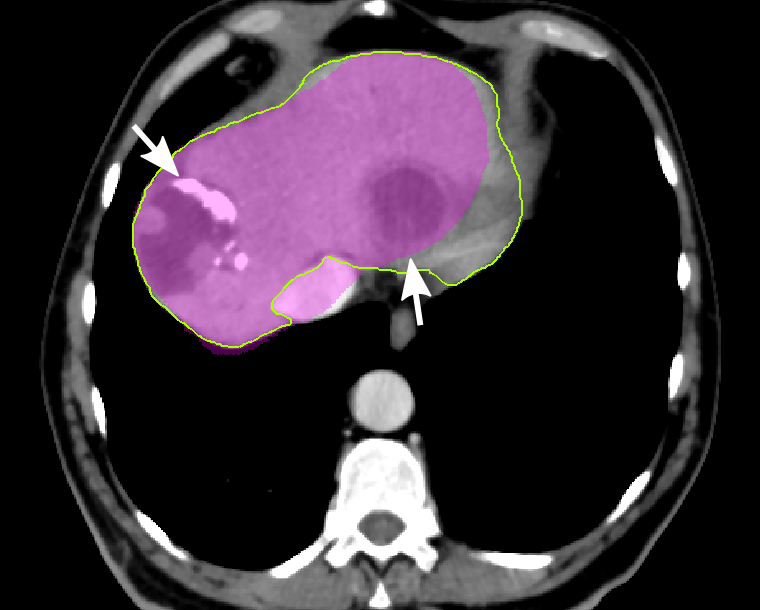}
				    \label{fig:qhull}
				\end{minipage}
			}
			\subfloat[]
			{
			    \begin{minipage}{0.24\textwidth}
				    \includegraphics[width=1.0\textwidth]{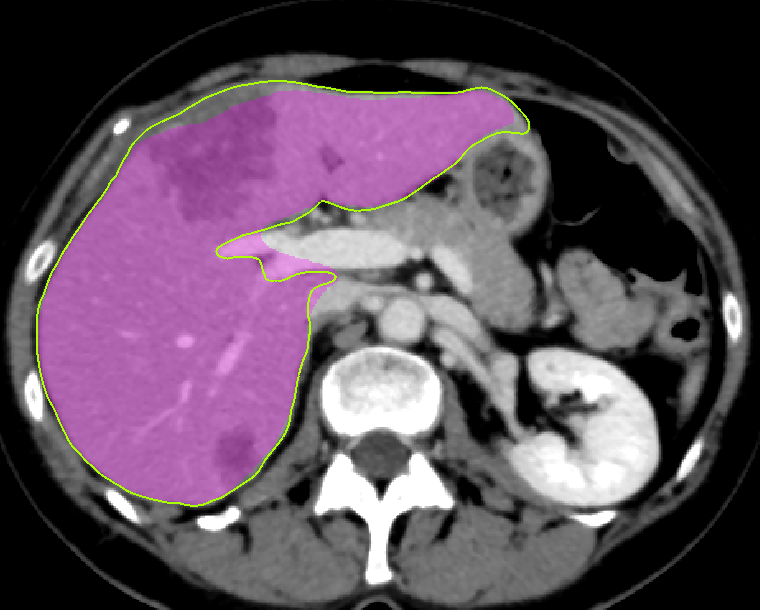}
				    \label{fig:qhull}
					
				    \includegraphics[width=1.0\textwidth]{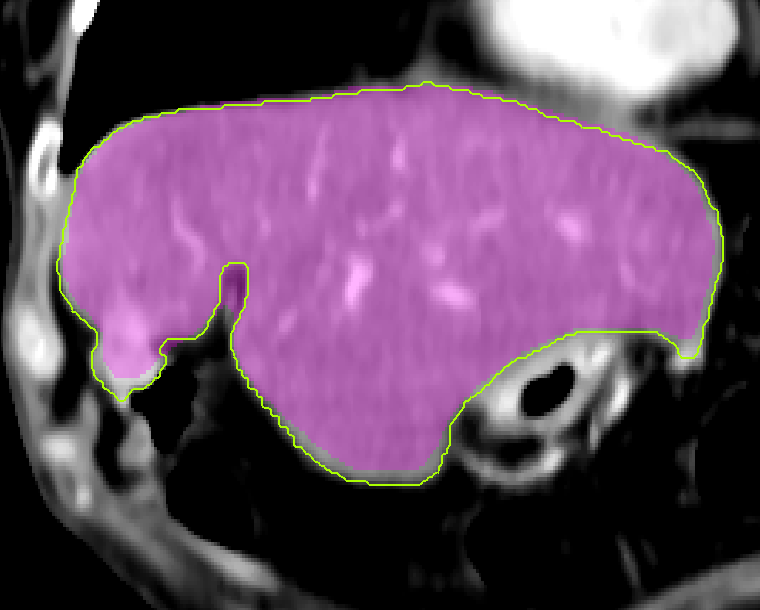}
				    \label{fig:qhull}
					
					\includegraphics[width=1.0\textwidth]{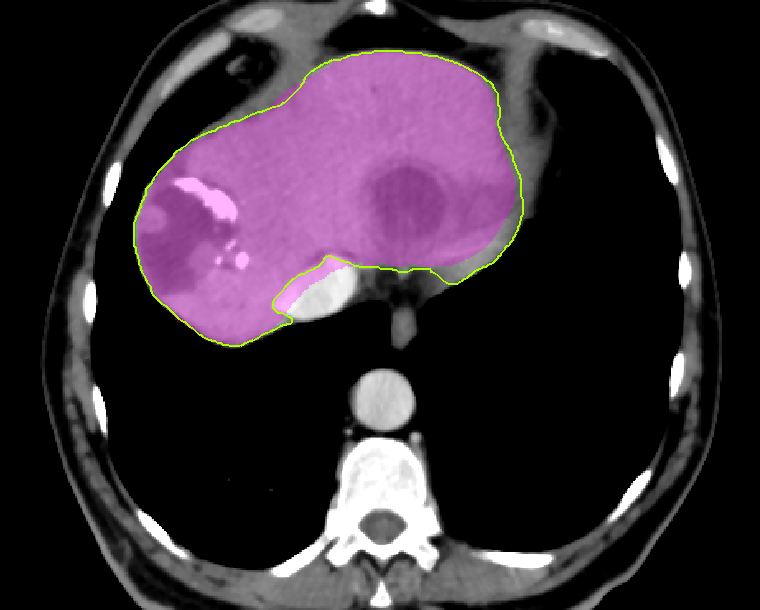}
				    \label{fig:qhull}
				\end{minipage}
			}
            \subfloat[]
			{
			    \begin{minipage}{0.24\textwidth}
				    \includegraphics[width=1.0\textwidth]{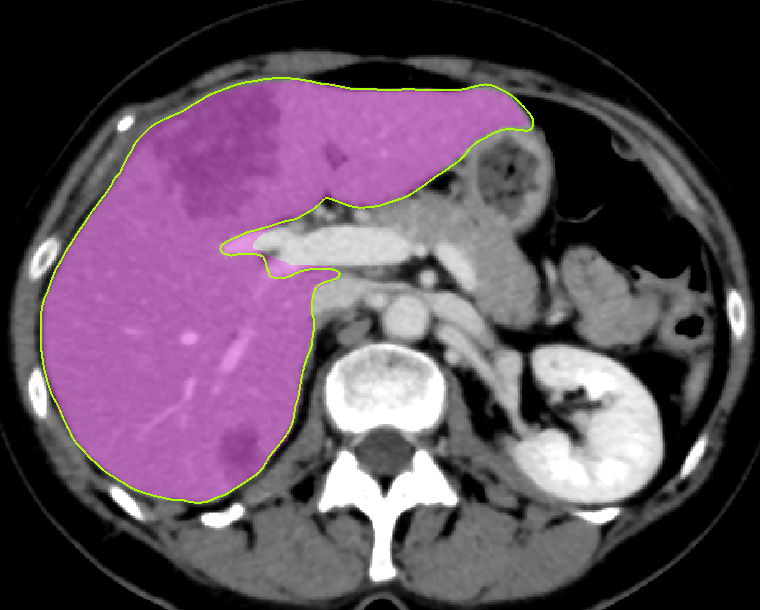}
				    \label{fig:qhull}
					
				    \includegraphics[width=1.0\textwidth]{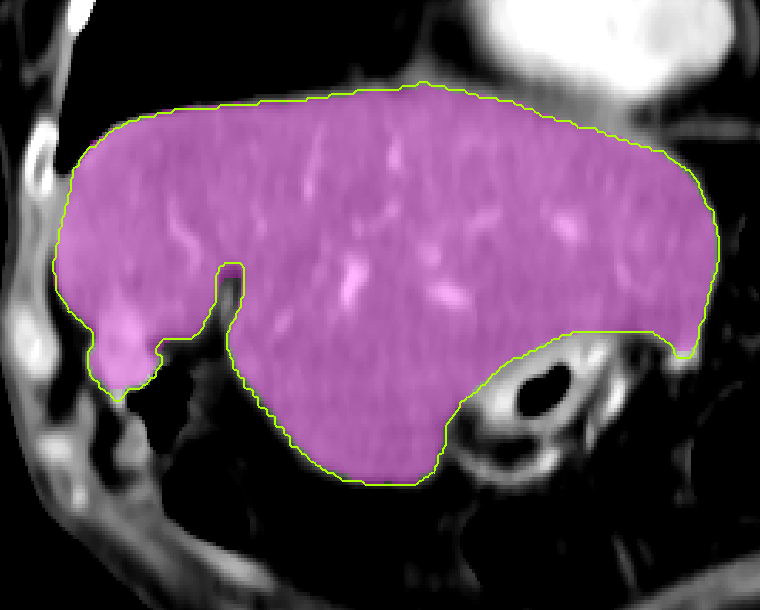}
				    \label{fig:qhull}
					
					\includegraphics[width=1.0\textwidth]{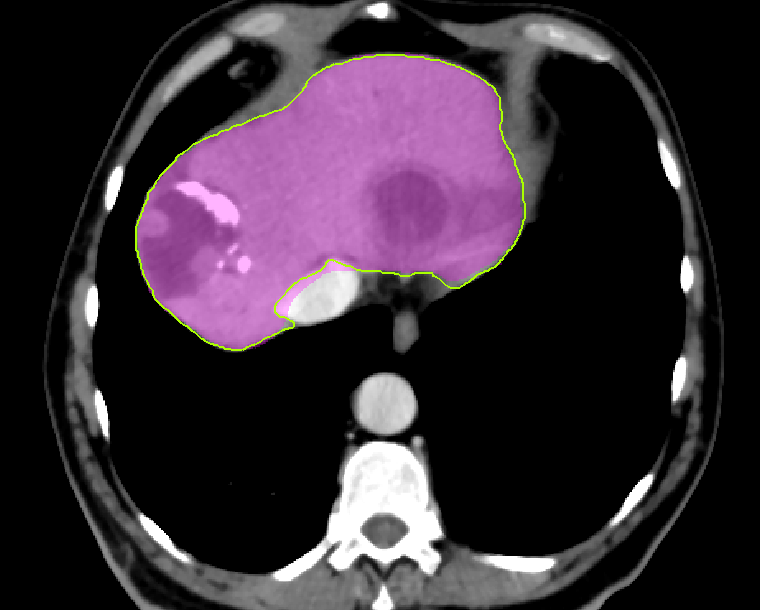}
				    \label{fig:qhull}
				\end{minipage}
			}
            \subfloat[]
			{
			    \begin{minipage}{0.24\textwidth}
				    \includegraphics[width=1.0\textwidth]{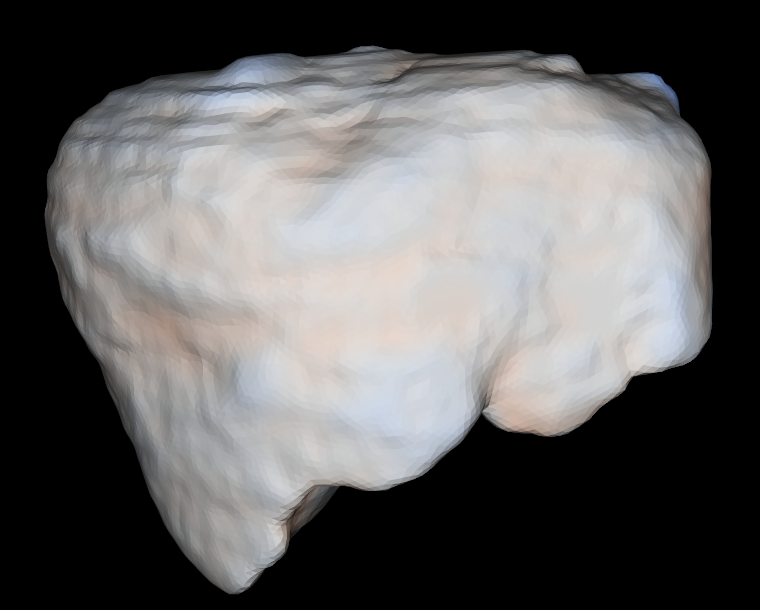}
				    \label{fig:qhull}
					
				    \includegraphics[width=1.0\textwidth]{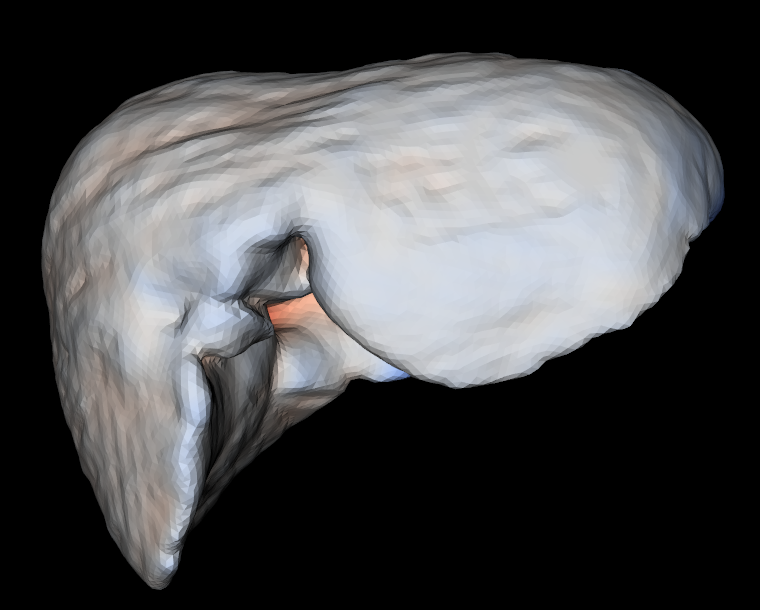}
				    \label{fig:qhull}
					
					\includegraphics[width=1.0\textwidth]{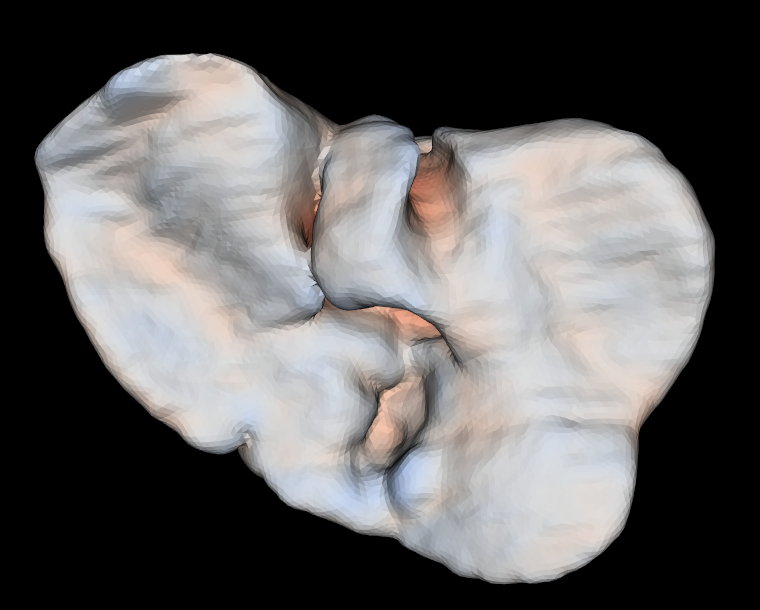}
				    \label{fig:qhull}
				\end{minipage}
			}
			\subfloat[]
			{
			    \begin{minipage}{0.1036\textwidth}
				    \vspace{7.95em}
				    \includegraphics[width=1.0\textwidth]{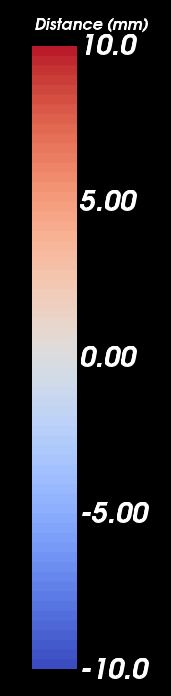}
				    \label{fig:qhull}
				\end{minipage}
			}

		\end{center}
	}	
		\caption{Results of liver segmentation by  
			  (a) the conventional MAS, (b) the LRMD-based MAS,  
	          and (c) the proposed LRTD-based MAS frameworks on three challenging pathological cases, 
			  consisting of liver tissue with hypodense tumor (1st row),
              hyperdense tumor (2nd row)			  
			  and hypo- and hyperdense tumors (3rd row),
			  indicated by white arrows.
			  The ground truths are delineated by green contours, 
			  while the segmentation results are shown as pink regions.		      			  
			  (d) The 3D visualization of 
			  average symmetric surface distance (ASD) errors of our method. 
              The red and blue regions indicate 
              over- and under-segmentation, respectively.			  
              (e) The distance to color bar.		  
			  }
		\label{fig:lrtd-segmentation}
\end{figure*}
%

Fig. \ref{fig:lrtd-segmentation} shows 
the visual results of liver segmentation generated by
the conventional MAS, the LRMD-based MAS,  
and our proposed LRTD-based MAS frameworks 
on three challenging pathological cases, 
consisting of liver tissue with hypodense tumor, 
hyperdense tumor, and hypo- and hyperdense tumors
(indicated by white arrows).
Each row shows one case. 
%
The first case is a very challenging one,
where the peripheral hypodense tumor
(indicated by the upper left white arrow) 
has very similar intensity values to that of the nearby muscle tissue.
We can see that the segmentation results of 
both the conventional MAS and the LRMD-based MAS
exclude part of the peripheral pathological areas,
thus resulting in under-segmentation.  
It is mainly due to the blurred boundaries 
between the peripheral hypodense tumor and 
the nearby muscle tissue,  
leading to inaccurate pairwise image registration
and label propagation 
in the peripheral pathological areas.
While our method accurately delineates 
the peripheral pathological areas,
and yields more accurate segmentation results.
%
%
This is mainly because
by generating more accurate tumor-free liver images,
our method can more effectively mitigate the
performance degradation of liver segmentation 
due to the presence of tumors,
compared to both the conventional MAS and the LRMD-based MAS.

In the second and the third cases,
the liver tumors are also located 
near the boundary of the liver tissue.
As expected, 
due to the inaccurate pairwise image registration 
in the peripheral pathological areas,
the segmentation result of the conventional MAS 
excludes a large portion of the peripheral liver tissue,
and thus under-segmenting the live tissue. 
On the other hand,
both the LRMD-based MAS and our method 
successfully delineate
the boundaries near the peripheral pathological areas,
and the segmentation results are much more accurate.
This is because in both methods,
the negative influence of these tumors
on liver segmentation results
can be largely mitigated by generating tumor-free liver images. 
However,
our method recovers more finer edge details 
and yields more accurate result than the LRMD-based MAS,
due to the use of more homogeneous and smooth tumor-free liver images,    
which lead to more accurate pairwise image registration
and label propagation.
These results demonstrate 
the strong robustness of our method
to both hypodense and hyperdense tumor contrast levels.

\subsubsection{ Quantitative Results }

Table \ref{table:Liver-Pathology_Results}
lists the quantitative comparative results of
liver segmentation with 
three different MAS frameworks,
i.e., the conventional MAS, the LRMD-based MAS, and the proposed LRTD-based MAS,
using the 3Dircadb1 database.
For each metric,
the mean and standard deviation of 
the overall datasets are given.
It can be observed that our LRTD-based MAS 
yields the best segmentation accuracy with the lowest variances
in terms of \responsess{all the evaluation metrics used}, 
demonstrating the robustness of our method
on a variety of test data.
Furthermore, 
our method outperforms the other two compared methods by a large margin, 
and statistically significant improvements ($p<0.05$) 
over both the conventional MAS and the LRMD-based MAS
were observed according to \responsess{all the evaluation metrics}.
%
%
%
\responsess{In particular, the mean $DC$, $JI$ and $ASD$ of our method are 
\SI{95.89}{\percent}, \SI{92.11}{\percent} and \SI{1.37}{\milli\metre}, respectively.}
For the conventional MAS,
the obtained mean $JI$ ($ASD$) is quite low (high),
with a value below \SI{85}{\percent} (above \SI{2.80}{\milli\metre}).
Compared to the LRMD-based MAS,
our method achieves an average improvement of 
\SI{1.58}{\percent} and \SI{18}{\percent}  
according to $JI$ and $ASD$, respectively.
The above results therefore clearly  
indicate that our proposed method
is significantly more accurate and robust
than both the conventional MAS and the LRMD-based MAS 
in the presence of major pathology.

\begin{table}[!t]
    \renewcommand{\arraystretch}{1.3}
    \caption{ \responsess{Comparative results of liver segmentation 
			  with three different MAS frameworks using the 3Dircadb1 database.	}		  
			}
    \label{table:Liver-Pathology_Results}
    \centering
	
\begin{tabular}{ p{12em} p{6.5em} p{6.5em} p{5.0em} }
\hline
Method & \responsess{DC} [\%] & JI [\%] & ASD [mm] \\
\hline
Conventional MAS       &   91.83$\pm$3.00$^\ast$  &  84.90$\pm$4.94$^\ast$  & 2.85$\pm$1.26$^\ast$  \\
LRMD-based MAS         &   95.02$\pm$1.28$^\ast$  &  90.53$\pm$2.30$^\ast$  &  1.67$\pm$0.47$^\ast$  \\
\textbf{Our LRTD-based MAS}  & \responsess{\textbf{95.89$\pm$0.51}}  & \textbf{92.11$\pm$0.95} & \textbf{1.37$\pm$0.21} \\

\hline
\end{tabular}

\begin{tabular}{ p{31em} }
{ \footnotesize For each metric, the mean and standard
                deviation of the overall datasets are given. 
              Bold values are the best result in that column. 
			  $\ast$ indicates a statistically significant difference 
			  between the marked result and the corresponding one of our method
			  at a significance level of $0.05$.
}
\end{tabular}

\end{table}

\responsess{
To validate the advantages of the LRTD for the image registrations, 
we compared the $NMI$ similarity measure~\citep{Studholme99} 
results of the three different MAS frameworks.
For $NMI \in \left[0, 1\right]$,
the larger the value is, 
the more accurate the image registrations will be.
Table \ref{table:Liver-Pathology_NMIResults}
summarizes the quantitative comparative results using the 3Dircadb1 database.
The mean and standard deviation of the overall datasets are given.
Our LRTD-based MAS achieves the highest mean $NMI$ of \SI{0.72},
while the mean $NMI$ of the conventional MAS is around \SI{0.6}..
The results clearly show the advantages of the LRTD 
for image registrations over both the LRMD and conventional method.
}

\begin{table}[!t]
    \renewcommand{\arraystretch}{1.3}
    \caption{ \responsess{Comparative results of the $NMI$ similarity measure with three different MAS frameworks using the 3Dircadb1 database.	}		  
			}
    \label{table:Liver-Pathology_NMIResults}
    \centering
	
\begin{tabular}{ p{12em} p{6.5em} p{6.5em} p{5.0em} }
\hline
\responsess{Method} & \responsess{$NMI$} \\
\hline
Conventional MAS       & 0.63$\pm$0.02 \\
LRMD-based MAS         & 0.67$\pm$0.02 \\
\textbf{Our LRTD-based MAS}  & \textbf{0.72$\pm$0.02} \\

\hline
\end{tabular}

\begin{tabular}{ p{31em} }
{ \footnotesize The mean and standard deviation of the overall datasets are given. 
              Bold values are the best result in that column. 
}
\end{tabular}

\end{table}

\begin{table}[!t]
    \renewcommand{\arraystretch}{1.3}
    \caption{ 
			  \responsess{Comparison of the proposed method and state-of-the-art automatic methods using the 3Dircadb1 database.} 
			}
    \label{table:Liver-Pathology_3Dircadb1}
    \centering
	
\begin{tabular}{ p{12.4em} p{8.0em} p{4.0em} p{5.0em} p{5.0em} }
\hline 
Method & Segmentation Framework & \responsess{DC} [\%] &JI [\%] & ASD [mm] \\
\hline  
\textit{Deep learning based methods} & & & \\

%
%
~\cite{Jiang19} & AHCNet &   94.50   &  89.57   &  N/A  \\
~\cite{Li18} & H-DenseUNet &   94.73   &  89.98$\pm$3.44 &  4.06$\pm$3.85 \\
~\cite{Kavur20} & Ensembles of DMs &  92.00    & 85.19   &  3.07  \\
~\cite{Lu17} & CNN+GC &  95.09    &  90.64$\pm$3.34  &  1.89$\pm$1.08  \\
~\cite{Christ17} &  Cascaded U-Net &  94.30    &  89.30  &  1.50\\
\hline

\textit{Model-based methods} & & & \\
~\cite{Lu18} & GC &  95.17    &  90.79$\pm$2.64  &  1.75$\pm$1.41  \\
~\cite{Erdt10} & ASM &  94.55   &  89.66$\pm$3.11  &  1.74$\pm$0.59  \\
~\cite{Esfandiarkhani17} & ASM &  94.52    & 89.61$\pm$2.45  &  1.66$\pm$0.48  \\
~\cite{Li15}  & GC &  95.21 &  90.85$\pm$1.44  &  1.55$\pm$0.39  \\
~\cite{shih16_PR} & ASM &  95.43    &  91.26$\pm$2.37  &  1.45$\pm$0.37  \\

\textbf{Our method} & MAS & \responsess{\textbf{95.89}}  & \textbf{92.11$\pm$0.95} & \textbf{1.37$\pm$0.21} \\ 
\hline
\end{tabular}

\begin{tabular}{ p{31em} }
{ \footnotesize Bold values are the best result in that column. 
			  ASM, GC and DM stand for active shape model, graph cut and deep learning model, respectively.
			  N/A stands for Not Available information.
}
\end{tabular}

\end{table}

\subsection{ Comparison with State-of-the-art Methods }
\label{subsec:sliver07-test} 

In this section,
we compared our proposed MAS-based liver
segmentation framework with state-of-the-art automatic methods 
using public databases
to evaluate its performance relative to the wider context of existing works.


We first compare our proposed liver segmentation framework with 
state-of-the-art automatic methods using the 3Dircadb1 database.
Table \ref{table:Liver-Pathology_3Dircadb1} 
outlines the quantitative results of liver
segmentation of all the compared methods,
including both deep learning based methods
and model-based methods.
As can be seen,
among all the 11 compared methods,
our method achieves the best performance,
\responsess{i.e., the largest (smallest) value of $DC$ and $JI$ ($ASD$).}
Also, our method yields very small variances
in terms of both $JI$ and $ASD$,
suggesting its robustness on a variety of datasets.
Among deep learning based methods,
the cascaded U-Net method proposed by~\cite{Christ17}
obtains the best accuracy,
\responsess{with the mean $DC$, $JI$ and $ASD$ of
\SI{94.30}{\percent}, \SI{89.30}{\percent} and \SI{1.50}{\milli\metre}, respectively.}
However,
the performance obtained by all other deep learning based methods
is inferior to that of all the model-based methods 
(including MAS, ASM and graph cut) in terms of $ASD$.
Although deep learning based methods 
are currently considered very good for medical image segmentation,
their performance is strongly dependent on
the availability of massive amounts of annotated training data.
Moreover, 
the lack of interpretability 
remains a major constraint to the adoption of deep learning based methods
in clinical applications,
where interpretability of the obtained results is of paramount importance.
Therefore, in clinical applications 
where the amount of annotated training data is low 
and a high level of trust is required (such as this study), 
model-based methods 
are still preferable to deep learning based methods. 

\begin{table}[!t]
    \renewcommand{\arraystretch}{1.3}
    \caption{ 
			  \responsess{Comparison of the proposed method and published state-of-the-art deep learning methods using the LiTS2017-Test database.} 
			}
    \label{table:Liver-Pathology_LiTS2017-Test}
    \centering
	
\begin{tabular}{ p{8.2em} p{16.0em} p{3.5em} p{3.5em} p{3.5em}   }
\hline 
\responsess{Method} & \responsess{Segmentation Framework}  & \responsess{DC} [\%] & \responsess{JI} [\%] & \responsess{ASD} [mm] \\
\hline  

\cite{Isensee21}  & \responsess{nnU-Net} &  \textbf{96.70}   & \textbf{93.60}  &  \textbf{0.97}\\

\cite{Liu18} & \responsess{3D Anisotropic Hybrid Network }   &  96.30   & 93.00  &   1.10\\
 
\cite{Yuan17}  & \responsess{Convolutional-Deconvolutional Neural Networks}   &  96.30 & 92.90  &  1.10\\
 
Tian  et al. (2017) & \responsess{FCN}   &  96.10 & 92.50  &   1.27\\

\cite{Li20} & \responsess{Bottleneck Feature Supervised U-Net}   &  96.10 & 92.50  &   1.42\\

\cite{Li18} & \responsess{H-DenseUNet}   &  96.10 & 92.60  &   1.45\\
 

Ben-Cohen  et al. (2017) & \responsess{FCN}   &  96.00 & 93.00  &   1.13\\

\cite{Chlebus17} & \responsess{2D U-Net}   &  96.00 & 92.30  &   1.15\\
 
Wu  et al. (2017) & \responsess{Cascaded FCNs}   &  95.90 & 92.10  &   1.31\\
 
Wang  et al. (2017) & \responsess{FCN}   &  95.80 & 92.00  &   1.37\\
 

\hline 

\textbf{Our method}  & \responsess{MAS}   &  95.30  & 91.00 & 1.43 \\

\hline
\end{tabular}

\begin{tabular}{ p{31em} }
{ \footnotesize 
              Bold values are the best result in that column. 
			  FCN stands for Fully Convolutional Network.
}
\end{tabular}

\end{table}

\responsess{
Furthermore, 
we participated the MICCAI 2017 Liver Tumor Segmentation (LiTS) challenge~\citep{LiTS17}, 
and compared the proposed liver segmentation framework 
with the top 10 published deep learning based methods.
The quantitative results are summarized in Table \ref{table:Liver-Pathology_LiTS2017-Test}.
The performance of our method
(team name: ivanshih~\footnote{\url{https://competitions.codalab.org/competitions/17094\#results}})
is comparable with that of state-of-the-art deep learning methods.
Specifically, the mean $DC$, $JI$ and $ASD$ of our method are 
\SI{95.30}{\percent}, \SI{91.00}{\percent} and \SI{1.43}{\milli\metre}, respectively.
Compared with the best-ranked method, i.e., nnU-Net~\citep{Isensee21},
the mean $DC$ and $JI$ of our method are only
\SI{1.40}{\percent} and \SI{2.60}{\percent} lower, respectively.
Nevertheless, all deep learning methods required substantially more training data
(LiTS2017-Train: 131) than ours (SLIVER07-Train: 20).
Furthermore, the performance of deep learning methods 
will degrade a lot if data augmentation was not applied~\citep{Rister20}.
}

All the above experimental results indicate that the performance of our method
is more accurate and robust than that of 
state-of-the-art methods,
including both deep learning based methods
and model-based methods.
Furthermore, 
the results demonstrate  
the strong robustness of our method 
against different tumor contrast levels.
Therefore,
the proposed MAS-based liver segmentation framework
can be utilized for accurate and robust liver segmentation
in the presence of major pathology.

%% file: chapters/conclusion.tex
\section{Conclusion}

In this paper, we propose 
a novel automatic method 
for accurate and robust 
pathological liver segmentation of CT images, 
by integrating the general $\star_{M}$-product based LRTD theory into 
the widely used MAS framework.
Our method significantly enhances the traditional MAS framework 
in three directions, 
by proposing a multi-slice LRTD scheme, 
an LRTD-based atlas construction method, 
and an LRTD-based MAS algorithm. 

To demonstrate the effectiveness of
our proposed segmentation framework,
we conducted extensive experiments    
using a total of \responsess{110} clinical CT scans
of pathological liver cases
from three publicly available databases, 
and our method yielded high performance.
All the experimental results 
indicate that:
\begin{itemize} 
\item The proposed multi-slice LRTD scheme is able to successfully
    recover the underlying low-rank structure embedded in 3D medical images.  
\item The proposed liver atlas construction method LRTD-PA 
	yields much more homogeneous and smooth tumor-free liver atlases than
	both the conventional PA and the LRMD-PA methods.    
\item The proposed LRTD-based MAS algorithm derives 
    very patient-specific liver atlases for each test image, 
    and achieves accurate pairwise 
	image registration and label propagation.	
%
\item The performance of our proposed segmentation framework
is superior to that of  
state-of-the-art methods,
including deep learning based methods
and model-based methods.
\end{itemize}
%

\responsess{
Despite encouraging results were obtained, 
some limitations still exist. Compared to deep learning and PA based methods, 
the computation time of our method is still high. 
It is mainly due to the computationally intensive nature 
of the pairwise non-rigid image registrations and the JLF algorithm. 
Although the training time of deep learning based methods 
can generally be more than several tens of hours, 
it takes a few minutes for them 
to perform inference on each test image~\citep{Li18}.
For PA-based methods, 
one non-rigid image registration between the PA and the test image 
was performed in testing stage; 
they are thus more efficient than our MAS-based method.
}

In future studies,
we plan to further improve its performance 
in two aspects:
%
(1) In the proposed multi-slice LRTD scheme, 
both the $\star_{M}$-product
and the tensor nuclear norm
depend on the transformation matrix $\mathbf{M}$ utilized.
We may further increase the accuracy of tensor decomposition 
by learning the optimal transformation matrix~\citep{Lu19_TRPCA}.
%
%
%
%
\responsess{
(2) 
	We will consider using deep learning-based registration methods,
    e.g., the VoxelMorph~\citep{Balakrishnan19},	
	to perform the image registrations
	to further decrease the computational time of our segmentation framework.  
}

%% file: chapters/appendix.tex
\section{ Tensor Preliminaries }
\label{sec_Appen:T-SVD}

\begin{equation}
\label{eq:LRSD-PCP1}
  \texttt{bcirc}(\bm{\mathcal{X}}) =
  \left[ \begin{array}{cccc}
 \mathbf{X}^{(1)}~&   \mathbf{X}^{(n_3)}~&    \ldots~& \mathbf{X}^{(2)}\\
 \mathbf{X}^{(2)}~&   \mathbf{X}^{(1)} ~&     \ldots~& \mathbf{X}^{(3)}\\
 \vdots~& \vdots~&    \ddots~& \vdots\\
 \mathbf{X}^{(n_3)}~& \mathbf{X}^{(n_3-1)}~&  \ldots~& \mathbf{X}^{(1)}
\end{array}\right ].
\end{equation}

\begin{equation}
\label{eq:LRSD-PCP2}
  \texttt{unfold}(\bm{\mathcal{X}}) =
  \left[ \begin{array}{c}
 \mathbf{X}^{(1)}\\
 \mathbf{X}^{(2)}\\
 \vdots\\
 \mathbf{X}^{(n_3)}
\end{array}\right ],\quad 
  \texttt{fold} \left( \texttt{unfold}(\bm{\mathcal{X}}) \right) = \bm{\mathcal{X}}.
\end{equation}

\begin{equation}
\label{eq:LRSD-PCP}
  \bar{\mathbf{X}} = \texttt{bdiag}(\bar{\bm{\mathcal{X}}}) =
  \left[ \begin{array}{cccc}
 \bar{\mathbf{X}}^{(1)}   ~&   ~&  ~& \\
 ~&\bar{\mathbf{X}}^{(2)} ~&  ~& \\
 ~& ~& \ddots~& \\
 ~& ~&       ~&\bar{\mathbf{X}}^{(n_3)}
\end{array}\right ].
\end{equation}

\begin{figure*}[!h]
	\begin{center}
		\includegraphics[width=1.0\textwidth]{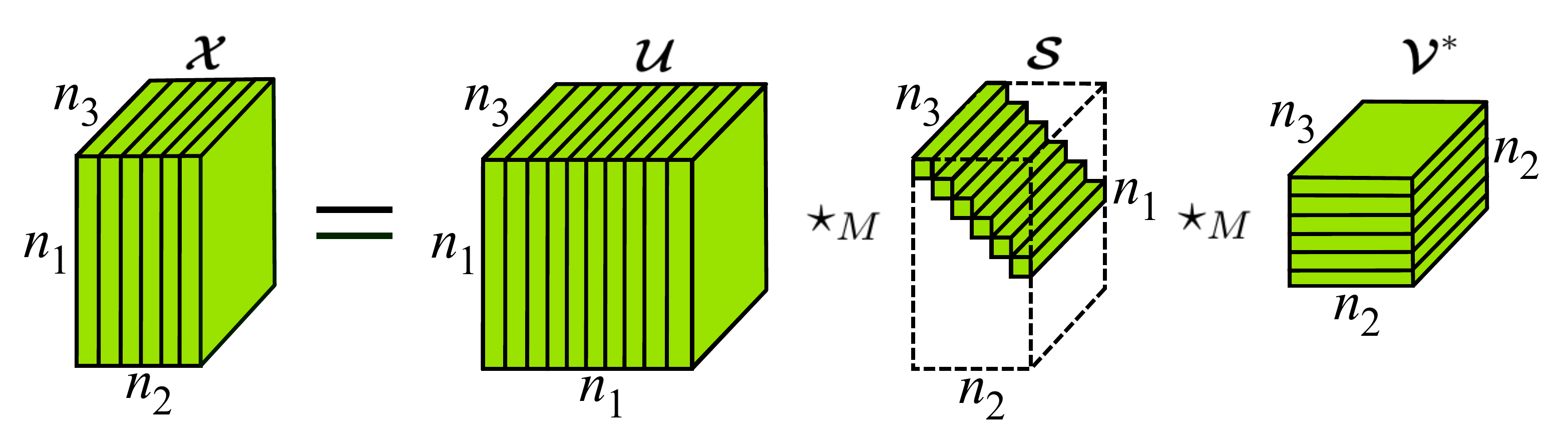}	
		\caption{An illustration of the t-SVD factorization of an $n_1 \times n_2 \times n_3$ tensor $\bm{\mathcal{X}}$.}
		\label{fig:tsvd}
	\end{center}
\end{figure*}
%

\begin{algorithm}[!h]
	\begin{center}	
    
    \begin{algorithmic}[0]
	

	\textbf{Input:}~tensor $\bm{\mathcal{X}} \in \mathbb{R}^{n_1 \times n_2 \times n_3}$ 
	                and invertible transformation matrix $\mathbf{M} \in \mathbb{R}^{n_3 \times n_3}$. \\ 
    
	\medskip
	 \textbf{Output:}~$\bm{\mathcal{U}} \in \mathbb{R}^{n_1 \times n_1 \times n_3}$, $\bm{\mathcal{S}} \in \mathbb{R}^{n_1 \times n_2 \times n_3}$, $\bm{\mathcal{V}} \in \mathbb{R}^{n_2 \times n_2 \times n_3}$.
	 \medskip

	\STATE{  \textbf{1.} Convert data into transform domain: $\bar{\bm{\mathcal{X}}} \leftarrow \mathbf{M}(\bm{\mathcal{X}})$ }.
	\medskip
	
	\STATE{ \textbf{2.} Compute matrix SVD for each frontal slice: \\ 
	\hspace{1.1em} $[\bar{\mathbf{U}}^{(k)}, \bar{\mathbf{S}}^{(k)}, \bar{\mathbf{V}}^{(k)}] \leftarrow $ \texttt{svd(}~$\bar{\bm{\mathcal{X}}}^{(k)}~\texttt{)}, k=1, ...,n_3$.}  
		
		
	
	\medskip
	\STATE{  \textbf{3.} Convert data back to spatial domain:\\
\hspace{1.1em}  $\bm{\mathcal{U}} \leftarrow \mathbf{M}^{-1}(\bar{\bm{\mathcal{U}}})$, $\bm{\mathcal{S}} \leftarrow \mathbf{M}^{-1}(\bar{\bm{\mathcal{S}}})$, $\bm{\mathcal{V}} \leftarrow \mathbf{M}^{-1}(\bar{\bm{\mathcal{V}}})$	}

    \end{algorithmic}

		\caption{The T-SVD Algorithm under Transformation Matrix $\mathbf{M}$}
        \label{algo:tsvd}
	\end{center}
\end{algorithm}
\begin{defs}[\textbf{Conjugate transpose}~\cite{Kilmer19}]
\label{shrinkage}
Let $\mathbf{M} \in \mathbb{R}^{n_3 \times n_3}$ be any invertible matrix.
The conjugate transpose of a tensor $\bm{\mathcal{X}} \in \mathbb{R}^{n_1 \times n_2 \times n_3}$ is the tensor $\bm{\mathcal{X}}^{*} \in \mathbb{R}^{n_2 \times n_1 \times n_3}$ satisfying 
$\left(\mathbf{M}({\bm{\mathcal{X}}^{*}})\right)^{(i)} = \left(\mathbf{M}({\bm{\mathcal{X}}})^{(i)} \right)^{*}$ for $i = 1,2,\cdots,n_3$.

\end{defs}

\begin{defs}[\textbf{Identity tensor}~\cite{Kilmer19}]
\label{shrinkage}
Let $\mathbf{M} \in \mathbb{R}^{n_3 \times n_3}$ be any invertible matrix. 
The tensor $\bm{\mathcal{I}} \in \mathbb{R}^{n \times n \times n_3}$
is an identity tensor if each frontal slice of $\bar{\bm{\mathcal{I}}} = \mathbf{M}(\bm{\mathcal{I}})$ is an $n \times n$ identity matrix $\mathbf{I}_{n}$.
 
\end{defs}

\begin{defs}[\textbf{Orthogonal tensor}~\cite{Kilmer19}] 
\label{shrinkage}
Let $\mathbf{M} \in \mathbb{R}^{n_3 \times n_3}$ be any invertible matrix.
A tensor $\bm{\mathcal{Q}} \in \mathbb{R}^{n \times n \times n_3}$ is orthogonal 
if it satisfies:
\begin{equation}
		\bm{\mathcal{Q}}^{*} \star_{M} \bm{\mathcal{Q}} = \bm{\mathcal{Q}} \star_{M} \bm{\mathcal{Q}}^{*} = \bm{\mathcal{I}}.  
\end{equation}
\end{defs}

\begin{defs}[\textbf{F-diagonal tensor}~\cite{Kilmer19}]
\label{shrinkage}
A tensor is f-diagonal if each of its frontal slice is diagonal.
 
\end{defs}

\section{ The Closed-Form Solutions to the TPCP Problem of LRTD-PA in Eq. \ref{eq:LRSD-TPCP2} via the ADMM Algorithm}
\label{sec_Appen:closed-form-ADMM}

The augmented Lagrangian function to be minimized for the TPCP problem in Eq. \ref{eq:LRSD-TPCP2} is given as follows (from now on the subscript $j$ is omitted to simplify notation):
\begin{equation}
\label{eq:Lagrangian}
		 L_{\mu}(\bm{\mathcal{L}}, \bm{\mathcal{E}}, \bm{\mathcal{Y}}) = 
    \|\bm{\mathcal{L}}\|_{*} + \lambda\|\bm{\mathcal{E}}\|_1 + 
				     \left \langle \bm{\mathcal{Y}}, \bm{\mathcal{X}} - \bm{\mathcal{L}} - \bm{\mathcal{E}} \right \rangle + \frac{\mu}{2} \|\bm{\mathcal{X}} - \bm{\mathcal{L}} - \bm{\mathcal{E}}\|_F^2,
\end{equation}
where $\bm{\mathcal{Y}}$ 
and $\mu > 0$ denote the Lagrange multiplier and penalty parameter, respectively. 
%
%
%
%
To solve the problem in Eq. \ref{eq:Lagrangian}, 
the ADMM algorithm first decomposes it into the following two minimization subproblems:
\begin{equation}
\label{eq:LRTD-A-two}
  \begin{split}
	& \bm{\mathcal{L}}_{k+1} = \operatorname*{arg\,min}_{\bm{\mathcal{L}}} ~L_{\mu_{k}}(\bm{\mathcal{L}}, \bm{\mathcal{E}}_k, \bm{\mathcal{Y}}_k), \\ 
  & \bm{\mathcal{E}}_{k+1} = \operatorname*{arg\,min}_{\bm{\mathcal{E}}} ~L_{\mu_{k}}(\bm{\mathcal{L}}_{k+1}, \bm{\mathcal{E}}, \bm{\mathcal{Y}}_k).	
  \end{split}
\end{equation}
Then $\bm{\mathcal{L}}$ and $\bm{\mathcal{E}}$ are updated alternately
by minimizing the augmented Lagrangian function with the other fixed.
Finally, the Lagrange multiplier $\bm{\mathcal{Y}}$   
is updated according to the following rule: 
%
%
\begin{equation}
\label{eq:LRSD-SM}
	\bm{\mathcal{Y}}_{k+1} = \bm{\mathcal{Y}}_{k} + \mu_{k}(\bm{\mathcal{X}} - \bm{\mathcal{L}}_{k+1} - \bm{\mathcal{E}}_{k+1}).	
\end{equation}
Furthermore, both minimization subproblems in Eq. \ref{eq:LRTD-A-two} 
have closed-form solutions.

\begin{algorithm}[!b] 
	\begin{center}	
    
    \begin{algorithmic}[0]
	

	\textbf{Input:}~Data tensor $\bm{\mathcal{X}}$, weighting parameter $\lambda$. \\ 
    
	\medskip
    \textbf{Output:}~$\bm{\mathcal{L}} = \bm{\mathcal{L}}_{k+1}, \bm{\mathcal{E}} = \bm{\mathcal{E}}_{k+1}$.\\
	
	\medskip
	\STATE{  \textbf{1.} Initialization: \\ 
	 \hspace{1.0em} $\bm{\mathcal{L}}_0 = \bm{\mathcal{E}}_0 = \bm{\mathcal{Y}}_0 = 0$, 
	$\mu_0 = 10^{-3}, \mu_{max} = 10^{10}, \rho = 1.1, 
	\varepsilon = 10^{-8}$, and $k = 0$. }
    \smallskip
	
	\STATE{ \textbf{2.} Solving the TPCP problem iteratively: }
	\WHILE{ \textit{Not Converged} }
	    \smallskip
		\STATE{ \textbf{2.1} Update $\bm{\mathcal{L}}$:}
        \STATE 	 \hspace{1.4em} $\bm{\mathcal{L}}_{k+1} \leftarrow \mathbf{D}_{\frac{1}{\mu_{k}}}(\bm{\mathcal{X}} - \bm{\mathcal{E}}_k + \frac{\bm{\mathcal{Y}}_k}{\mu_k})$.		
		
		\medskip
		\STATE{ \textbf{2.2} Update $\bm{\mathcal{E}}$:}
        \STATE 	 \hspace{1.4em} $\bm{\mathcal{E}}_{k+1} \leftarrow \mathbf{S}_{\frac{\lambda}{\mu_k}} (\bm{\mathcal{X}} - \bm{\mathcal{L}}_{k+1} + \frac{\bm{\mathcal{Y}}_k}{\mu_k})$.

        \medskip
		\STATE{ \textbf{2.3} Check the convergence conditions:} 
		\IF{ $\| \bm{\mathcal{L}}_{k+1} - \bm{\mathcal{L}}_{k} \|_{\infty}  < \varepsilon$ \AND $\| \bm{\mathcal{E}}_{k+1}- \bm{\mathcal{E}}_{k} \|_{\infty}  < \varepsilon$ \AND $\| \bm{\mathcal{X}} - \bm{\mathcal{L}}_{k+1} - \bm{\mathcal{E}}_{k+1} \|_{\infty}  < \varepsilon$ }
		    \STATE \textbf{break}. 
		\ENDIF

		\medskip
		\STATE{ \textbf{2.4} Update $\bm{\mathcal{Y}}$:}  
        \STATE 	 \hspace{1.4em} $\bm{\mathcal{Y}}_{k+1} \leftarrow \bm{\mathcal{Y}}_{k} + \mu_k(\bm{\mathcal{X}} - \bm{\mathcal{L}}_{k+1} - \bm{\mathcal{E}}_{k+1})$.
	    
		\medskip
		\STATE{ \textbf{2.5} Update $\mathbf{\mu}$:} 
		\STATE  \hspace{1.4em} $\mu_{k+1} \leftarrow \texttt{min(}~\rho\mu_{k}, \mu_{max}~\texttt{)}$.

		\medskip
		\STATE \textbf{2.6} $k \leftarrow k+1$.
		
		\medskip
    \ENDWHILE

    \end{algorithmic}

		\caption{The ADMM Algorithm for Solving the TPCP Problem in Eq. \ref{eq:LRSD-TPCP2}}
        \label{algo:ADMM}
	\end{center}
\end{algorithm}
%

\begin{algorithm}[!t] 
	\begin{center}	
    
    \begin{algorithmic}[0]
	

	\textbf{Input:}~Aligned training image tensors of cluster $c$: $\{\bm{\mathcal{D}}_i~|~i = 1, . . ., N_{c} \}$, weighting parameter $\lambda$, and the segment length $K$. \\ 
    
	\medskip
	\textbf{Output:}~$\hat{\bm{\mathcal{L}}}, \hat{\bm{\mathcal{E}}}$. 
	\medskip
	    \STATE{ \textbf{1.} Partition $\bm{\mathcal{D}}_i$ into $N_{s}$ image segments consisting of multiple consecutive image slices of length $K$: $[\bm{\mathcal{D}}_{i1},\bm{\mathcal{D}}_{i2}, \cdot\cdot\cdot, \bm{\mathcal{D}}_{iN_{s}}]~\leftarrow~\bm{\mathcal{D}}_i, i = 1, . . ., N_{c}$}.

	
    \medskip
	\STATE{ \textbf{2.} Perform the LRTD on each image segment tensor: }
	\FOR{$j=1$ \TO $N_{s}$}
	    \STATE{ \textbf{2.1} $\bm{\mathcal{X}}_{j}~\leftarrow~$Construct an image repository tensor for segment $j$ by using the corresponding training segments $\{\bm{\mathcal{D}}_{ij} ~|~i = 1, . . ., N_{c} \}$.}
		
	\medskip
		\STATE{ \textbf{2.2} Use the ADMM Algorithm (Algorithm~\ref{algo:ADMM}) to perform LRTD: \\
		 \hspace{1.2em} $(\hat{\bm{\mathcal{L}}_{j}}, \hat{\bm{\mathcal{E}}_{j}})~\leftarrow~\texttt{ADMM(}\bm{\mathcal{X}}_{j}, \lambda\texttt{)}$ }.
	\ENDFOR

    \medskip
	\STATE{ \textbf{3.} $(\hat{\bm{\mathcal{L}}}, \hat{\bm{\mathcal{E}}})~\leftarrow~$Stack $\{\hat{\bm{\mathcal{L}}_{j}}~|~j = 1, . . ., N_{s} \}$, $\{\hat{\bm{\mathcal{E}}_{j}}~|~j = 1, . . ., N_{s} \}$ } frontal-slice-wisely.

    \end{algorithmic}

		\caption{Optimization Procedure of the Multi-Slice LRTD Scheme}
        \label{algo:ADMM2}
	\end{center}
\end{algorithm}
%

(i) \emph{$\bm{\mathcal{L}}$ minimization subproblem}:
\begin{thm}[~\citet{Lu20}]
\label{SVT}
Given a tensor $\bm{\mathcal{W}} \in \mathbb{R}^{n_1 \times n_2 \times n_3}$ and $\tau > 0$,
the optimal solution to the following 
minimization problem is given by:
\begin{equation}
		\mathbf{D}_{\tau}(\bm{\mathcal{W}}) = \operatorname*{arg\,min}_{\bm{\mathcal{X}}} ~\tau \|\bm{\mathcal{X}}\|_{*} + \frac{1}{2}\| \bm{\mathcal{X}} - \bm{\mathcal{W}} \|_F^2, 
\end{equation}
%
where $\mathbf{D}_{\tau}$ is the tensor singular value thresholding (t-SVT) operator~\citep{Lu20} defined as:
$\mathbf{D}_{\tau}(\bm{\mathcal{W}}) = \bm{\mathcal{U}} \star_{M} \bm{\mathcal{S}}_{\tau} \star_{M} \bm{\mathcal{V}}^{*}$,  
where $\bm{\mathcal{U}}\star_{M}\bm{\mathcal{S}}\star_{M}\bm{\mathcal{V}}^{*} = \bm{\mathcal{W}}$ 
is the t-SVD of $\bm{\mathcal{W}}$, 
$\bm{\mathcal{S}}_{\tau} = \mathbf{M}^{-1}\left(\mathbf{S}_{\tau}( \bar{\bm{\mathcal{S}}} )\right)$, 
and $\mathbf{S}_{\tau}( \bar{s}_{ijk} ) = \textnormal{max}(|\bar{s}_{ijk}| - \tau, 0) \cdot \textnormal{sgn}(\bar{s}_{ijk})$ is the shrinkage operator applied on $\bar{\bm{\mathcal{S}}}$ element-wisely, 
where sgn($\cdot$) is the sign function.
\end{thm}

Given the other variable fixed, 
the closed-form solution for 
$\bm{\mathcal{L}}$ subproblem in Eq. \ref{eq:LRTD-A-two} 
can be obtained
as follows according to Theorem \ref{SVT}:
 
\begin{equation}
\label{eq:LRSD-A-solution}
	\begin{split}
		&\bm{\mathcal{L}}_{k+1} = \operatorname*{arg\,min}_{\bm{\mathcal{L}}} ~L_{\mu_{k}}(\bm{\mathcal{L}}, \bm{\mathcal{E}}_k, \mathbf{Y}_k)\\
		&=\operatorname*{arg\,min}_{\bm{\mathcal{L}}} ~\|\bm{\mathcal{L}}\|_{*} + \frac{\mu_{k}}{2} \|\bm{\mathcal{X}} - \bm{\mathcal{L}} - \bm{\mathcal{E}}_k + \frac{\bm{\mathcal{Y}}_k}{\mu_k}\|_F^2\\ 
				    &= \mathbf{D}_{\frac{1}{\mu_{k}}}(\bm{\mathcal{X}} - \bm{\mathcal{E}}_k + \frac{\bm{\mathcal{Y}}_k}{\mu_k}).
	\end{split}
\end{equation}

(ii) \emph{$\bm{\mathcal{E}}$ minimization subproblem}:
\begin{thm}[~\citet{Hale08}] 
\label{shrinkage}
Given a tensor $\bm{\mathcal{W}} \in \mathbb{R}^{n_1 \times n_2 \times n_3}$ and $\tau > 0$,
the optimal solution to the following 
minimization problem is given by: 
\begin{equation}
		\mathbf{S}_{\tau}(\bm{\mathcal{W}}) = \operatorname*{arg\,min}_{\bm{\mathcal{X}}} ~\tau \|\bm{\mathcal{X}}\|_1 + \frac{1}{2}\| \bm{\mathcal{X}} - \bm{\mathcal{W}} \|_F^2,
\end{equation}
where $\mathbf{S}_{\tau}$ is the shrinkage operator.
\end{thm}

Similarly, 
the closed-form solution for 
$\bm{\mathcal{E}}$ subproblem in Eq. \ref{eq:LRTD-A-two} 
can be written
as follows according to Theorem \ref{shrinkage}:
\begin{equation}
\label{eq:MLR-SSC}
	\begin{split}
	    &\bm{\mathcal{E}}_{k+1} = \operatorname*{arg\,min}_{\bm{\mathcal{E}}} ~L_{\mu_{k}}(\bm{\mathcal{L}}_{k+1}, \bm{\mathcal{E}}, \bm{\mathcal{Y}}_k)\\
		&=\operatorname*{arg\,min}_{\bm{\mathcal{E}}} ~\lambda\|\bm{\mathcal{E}}\|_1 + \frac{\mu_{k}}{2} \|\bm{\mathcal{X}} - \bm{\mathcal{L}}_{k+1} - \bm{\mathcal{E}} + \frac{\bm{\mathcal{Y}}_k}{\mu_k}\|_F^2\\ 
				    &= \mathbf{S}_{\frac{\lambda}{\mu_{k}}}(\bm{\mathcal{X}} - \bm{\mathcal{L}}_{k+1} + \frac{\bm{\mathcal{Y}}_k}{\mu_k}).
	\end{split}
\end{equation}